\documentstyle[12pt]{article}
\begin{document}
\def\theequation{\thesection.\arabic{equation}}
\newenvironment{proof}{\noindent {\em Proof. }}{\hfill$\Box$.\\}
\newtheorem{theo}{Theorem}[section]
\newtheorem{prop}[theo]{Proposition}
\newtheorem{lemm}[theo]{Lemma}
\newtheorem{coro}[theo]{Corollary}
\newtheorem{defi}[theo]{Definition}
\newtheorem{rema}[theo]{Remark}
\newcommand{\sectio}[1]{\section{#1}\setcounter{equation}{0}}
\newcommand{\e}{\label}
\newcommand{\r}[1]{(\ref{#1})}
\newcommand{\re}{\ref}
\newcommand{\k}{\ldots}
\newcommand{\api}{{\cal A}}
\newcommand{\bpi}{{\cal B}}
\newcommand{\cpi}{{\cal C}}
\newcommand{\lpi}{{\cal L}}
\newcommand{\ppi}{{\cal P}}
\newcommand{\spi}{{\cal I}}
\newcommand{\nb}{\mbox{{\bf N}}{}}
\newcommand{\rb}{\mbox{{\bf R}}{}}
\newcommand{\cb}{\mbox{{\bf C}}{}}
\newcommand{\jb}{\mbox{{\bf 1}}{}}
\newcommand{\al}{\alpha}
\newcommand{\ga}{\gamma}
\newcommand{\del}{\delta}
\newcommand{\eps}{\epsilon}
\newcommand{\es}{s}
\newcommand{\lam}{\lambda}
\newcommand{\si}{\sigma}
\newcommand{\de}{\Delta}
\newcommand{\la}{\Lambda}
\newcommand{\wb}{\bar w}
\newcommand{\g}{\Gamma}
\newcommand{\gd}{\dot\Gamma}
\newcommand{\poi}{Poincar\'e\ }
\newcommand{\qpg}{quantum \poi  group}
\newcommand{\qig}{quantum inhomogeneous group}
\newcommand{\be}{\begin{equation}}
\newcommand{\ee}{\end{equation}}
\newcommand{\bt}{\begin{theo}}
\newcommand{\et}{\end{theo}}
\newcommand{\bp}{\begin{prop}}
\newcommand{\ep}{\end{prop}}
\newcommand{\bl}{\begin{lemm}}
\newcommand{\el}{\end{lemm}}
\newcommand{\bc}{\begin{coro}}
\newcommand{\ec}{\end{coro}}
\newcommand{\bde}{\begin{defi}}
\newcommand{\ede}{\end{defi}}
\newcommand{\br}{\begin{rema}}
\newcommand{\er}{\end{rema}}
\newcommand{\bd}{\begin{proof}}
\newcommand{\ed}{\end{proof}}
\newcommand{\ba}{\begin{array}}
\newcommand{\ea}{\end{array}}
\newcommand{\btr}{\begin{trivlist}}
\newcommand{\etr}{\end{trivlist}}
\newcommand{\lra}{\longrightarrow}
\newcommand{\ot}{\otimes}
\newcommand{\op}{\oplus}
\newcommand{\tp}{\mbox{\raisebox{0.55mm}{\small{$\bigcirc$}}%
                 \hspace{-1.9ex}\small{$\top$}}}
\newcommand{\ti}{\mbox{\raisebox{0.55mm}{\small{$\bigcirc$}}%
                 \hspace{-1.9ex}\raisebox{0.65mm}{\small{$\bot$}}}}         
\newcommand{\tu}{\mbox{\hspace{-0.4ex}%
                 \raisebox{0.55mm}{\scriptsize{$\bigcirc$}}%
                 \hspace{-1.45ex}\scriptsize{$\top$}\hspace{0.2ex}}}
\newcommand{\sd}{\rhd\mbox{\hspace{-2ex}}<}		 
\newcommand{\spa}{\mbox{{\rm{span}}}}                 
\newcommand{\po}{\mbox{{\rm{Poly}}}}
\newcommand{\id}{\mbox{{\rm{id}}}}
\newcommand{\mor}{\mbox{{\rm{Mor}}}}
\newcommand{\te}{\tilde}
\newcommand{\tb}{\te{\cal B}}
\newcommand{\iffi}{\Leftrightarrow}
\newcommand{\mod}{\mbox{{\rm{mod}}}}
\newcommand{\im}{\mbox{{\rm{im}}}}
\newcommand{\Sp}{\mbox{{\rm{Sp}}}}
\newcommand{\tr}{\mbox{{\rm{tr}}}}
\newcommand{\Irr}{\mbox{{\rm{Irr}}}}
\newcommand{\Rep}{\mbox{{\rm{Rep}}}}
\newcommand{\ov}{\overline}
\newcommand{\ite}{\item[]}
\newcommand{\hs}{\hspace{0.2ex}}
\newcommand{\pri}{{}^{\prime}}
\newcommand{\itemite}{\item[]\hspace{3ex}}
\newcommand{\Ren}{\mbox{{\rm{Re}}}}
\newcommand{\Imn}{\mbox{{\rm{Im}}}}
\title{On the structure of inhomogeneous quantum groups}
\author{P. Podle\'s${}^1$\thanks{On leave from 
Department of Mathematical Methods in Physics,
Faculty of Physics, University of Warsaw,
Ho\.za 74, 00--682 Warszawa, Poland}
{}\thanks{This research was supported in part
by NSF grant DMS92--43893 and in part by Polish KBN grant 
No 2 P301 020 07}
and S. L. Woronowicz${}^2$\thanks{This research was supported by Polish KBN
grant No 2 P301 020 07}\\
${}^1$ Department of Mathematics, University of California\\
Berkeley CA 94720, USA\\
${}^2$ Department of Mathematical Methods in Physics\\
Faculty of Physics, University of Warsaw\\
Ho\.za 74, 00--682 Warszawa, Poland}
\date{}
\maketitle
\begin{abstract}
We investigate inhomogeneous quantum groups $G$ 
built from a quantum group $H$ and translations. The corresponding
commutation relations contain inhomogeneous terms. Under certain conditions
(which are satisfied in our study of quantum Poincar\'e groups \cite{QP})
we prove that our construction has correct `size', find the 
$R$-matrices and the analogues of Minkowski space for $G$.
\end{abstract}
\setcounter{section}{-1}
\sectio{Introduction}

Inhomogeneous quantum groups,
their homogeneous spaces and corresponding R-matrices
 were studied by many authors (cf e.g. \cite{Cel},
\cite{L}, \cite{D}, \cite{S}, \cite{O}, \cite{M}, \cite{ChD}). Here we
propose a general construction which covers the examples \cite{L}, \cite{ChD}
and is suitable for our study of quantum Poincar\'e groups 
(without dilatations) \cite{QP}. We work in the framework of 
Hopf algebras treated as algebras of functions on quantum groups. 
In Section~1 we assume that $G$ is an inhomogeneous quantum group built from
a quantum group $H$ and translations described by the elements $p_i$ 
corresponding to an irreducible representation $\la$ of $H$. The commutation 
relations can contain inhomogeneous terms. It turns out that the leading terms 
in these relations are governed by the structure of certain bicovariant 
bimodule of $H$. In particular, the leading terms in relations among $p_i$
must correspond to the eigenvalue $-1$ of the corresponding R-matrix $R$ (cf 
\cite{PW}, \cite{M}). In Section 2 we add the condition that all the 
representations of $H$ are completely reducible (which is the case in 
\cite{QP} when $H$ is a quantum Lorentz group \cite{WZ}) and find the
commutation relations between functions on $H$ and $p_i$. In Section~3 we
assume that $R^2=\jb$ 
(or $(R-Q\jb)(R+\jb)=0$ where $Q\neq0$ is not a root of unity)
and that we have as many quadratic relations among 
$p_i$ as it is allowed by the eigenvalue $-1$ property (so, if there would be
no inhomogeneous terms, $p_i$ would be $R$-symmetric). That is the simplest
case which is sufficient in \cite{QP}. We find the exact form of commutation 
relations and the necessary and sufficient
conditions for the corresponding coefficients. If they are 
fulfilled, there are no relations of higher order and our construction has the 
same `size' as in the absence of inhomogeneous terms. The R-matrices
for the fundamental representation of $G$ are classified. In Section 4
we consider the 
$*$-structure and isomorphisms among our objects. In Section~5
we prove that (under some conditions which are fulfilled in \cite{QP}) each
$G$ possesses exactly one analogue of Minkowski space.
Inhomogeneous Poisson groups are considered in \cite{ZAK}.

For the simplicity of calculations, the small Latin indices $a,b,c,d,\ldots,$
belong to a finite set $\spi$ in Sections~1--5. 
We sum over
repeated indices which are not taken in brackets (Einstein's convention). The
number of elements in a set $B$ is $\#B$
or $|B|$. We work over the field $\cb$.
Unit matrix with dimension $N$ is
denoted by $\jb_N$.
If $V, W$ are vector spaces then $\tau_{VW}:V\ot W\lra W\ot V$ is given by
$\tau_{VW}(x\ot y) = y\ot x$, $x\in V$, $y\in W$. We often write $\tau$
instead of $\tau_{VW}$. If $\api$ is a linear space and $v,v'\in M_N(\api)$,
$N\in\nb$,
then $v\ti v'\in M_N(\api\ot\api)$ is defined by $(v\ti v')_{ij}=v_{ik}\ot
{v\pri}_{kj}$, $i,j=1,\ldots,N$ (Einstein's convention!). If moreover $\api$
is an algebra, $v\in M_N(\api)$, $w\in M_K(\api)$,then 
$v\tp w\in M_{NK}(\api)$ is defined by $(v\tp w)_{ij,kl}=v_{ik}w_{jl}$, 
$i,k=1,\k,N$, $j,l=1,\k,K$. We use the
abbreviation $v^{\tu n}$ for $v\tp\ldots\tp v$ ($n$ times). If $\api=\cb$ we
may write $\ot$ instead of $\tp$.
If also $v'\in M_N(\api), w'\in M_K(\api)$ 
then
$(v\tp w)\ti(v'\tp w')=(v\ti v')\tp(w\ti w')$ (see (2.18) of \cite{W1}). 
If $\api$ is a ${}^*$-algebra then the conjugate of $v$ is defined as 
$\bar v\in M_N(\api)$, where $\bar v_{ij}={v_{ij}}^*$, $i,j=1,\k,N$. 

Throughout the paper quantum groups $G$ are abstract objects described by the
corresponding Hopf (${}^*$-) algebras $\po(G)=(\api,\de)$. We denote by
$\de,\eps,S$ the comultiplication, counit and coinverse of $\po(G)$. 
We always assume that $S$ is invertible. 
We say 
that $v$ is a representation of $G$ 
(i.e. $v\in\Rep\ G$) if $v\in M_N(\api),$ $N\in \nb,$ and $\de
v_{ij}=v_{ik}\ot v_{kj}$, $\eps(v_{ij})=\delta_{ij}$, $i,j=1,2,\ldots,N$.
Then $\dim\ v=N$. The conjugate of a representation and tensor product
of representations are also representations. 
Matrix elements of representations of $G$
span $\api$ as a linear space. The set of nonequivalent
irreducible representations of $G$
is denoted by $\Irr\ G$.
If $v,w\in\Rep\ G$ then we say that 
$A\in M_{\dim w\times\dim v}(\cb)$ intertwines $v$ with $w$ (i.e.
$A\in\mor(v,w)$) if $Av=wA$. We use the following notations:
\[ a*\rho=(\rho\ot\id)\de a,\quad \rho*a=(\id\ot\rho)\de a, \quad
\rho*\rho'=(\rho\ot\rho')\de \]
for $a\in\api$, $\rho,\rho'\in\api'$. Let us recall the following
well--known

\bp\e{p0.1} Let $\api$ be an augmented algebra
(i.e. a unital algebra endowed with a unital homomorphism
$\epsilon:\api\lra\cb$)
 and $\omega_i$, $i=1,2,\k,M$, and 
$\eta_j$, $j=1,2,\k,N$, be two free bases of the left $\api$-module $\g$. 
Then $M=N$.
\ep

\bd One has $\omega_i=a_{ij}\eta_j$, $\eta_j=b_{ji}\omega_i$ for some 
$a_{ij},
b_{ji}\in \api$. It follows $a_{ij}b_{jk}=\delta_{ik}I$, $b_{li}a_{ij}=
\delta_{lj}I$, $k=1,\k,M$, $l=1,\k,N$. Applying $\eps$, one gets 
$\eps(a)\eps(b)=\jb_M$, $\eps(b)\eps(a)=\jb_N$, hence $\eps(a)$ 
is invertible, $M=N$. \ed

Therefore the dimension $\dim_\api$ of a free left $\api$-module (where
$\api$ is a Hopf algebra)
is well defined. We can use the above notions and facts 
(if applicable) also 
for general (${}^*$-) bialgebras (without $S$) 
and not necessarily quadratic matrices.

\sectio{ Inhomogeneous quantum groups}

In this Section we define inhomogeneous quantum groups and study leading
terms in their commutation relations using the theory of bicovariant
bimodules \cite{W2}. The importance of left covariant bimodule structure 
in investigation of inhomogeneous quantum groups was first noticed in \cite{S}.
 
Let us assume that $\po(H)=(\api,\de)$ is a Hopf algebra with a distinguished
irreducible
representation $\la=(\la_{rs})_{r,s\in\spi}$ of $H$, $|\spi|<\infty$.
We set  $\jb=\jb_{|\spi|}$. We shall
consider bialgebras $\po(G)=(\bpi,\de)$ such that:
\btr
\item[1.] $\bpi$ is generated (as algebra) by $\api$ and
the elements $p_s$, $s\in\spi$.
\item[2.] $\api$ is a sub--bialgebra of $\bpi$.
\item[3.] $\ppi=\left(\ba{cc} \la&p\\0&I \ea\right)$ is a representation of
$G$. 
\item[4.]  There exists $i\in\spi$ such that $p_i\not\in\api$.
\item[5.] $\g\api\subset\g$ where $\g=\api\cdot X+\api$, $X=\spa\{p_i,\ \
i\in\spi\}$. \etr

Due to 5., $\g$ is an $\api$-bimodule. In virtue of 2.-3., $\de
\api\subset \api\ot \api$,
\be \de p=\la\ti p+p\ti I,\e{2.0}\ee
hence $\de\g\subset \api\ot\g+\g\ot\api$. We define bimodule $\gd=\g/{\api}$ 
by $a\dot\omega=\dot{a\omega}$, $\dot\omega a=\dot{\omega a}$ where  
$\dot\omega$ is the element of $\gd$ corresponding to $\omega\in\g$,
$a\in\api$. We see that $\de$ induces a linear mapping 
\[ \dot\de:\gd\lra
(\api\ot\g+\g\ot\api)/(\api\ot\api)\approx(\api\ot\gd)\op(\gd\ot\api). \]
We get the decomposition $\dot\de=\de_L+\de_R$, $\de_L:\gd\lra\api\ot\gd$,
$\de_R:\gd\lra\gd\ot\api$. In particular,
$\de_L\dot{p_s}=\la_{st}\ot\dot{p_t}$, $\de_R\dot{p_s}=\dot{p_s}\ot I$. Using
the properties of $\de$, one can easily check that $(\gd,\de_L,\de_R)$ is a
bicovariant bimodule (cf \cite{W2}, Definition 2.3
and a similar argument in the proof of Theorem 5 of \cite{B}). 
We notice that 
$\dot{p_s}$
($s\in\spi$) are elements in the
set $\dot\g_{inv}$ of right--invariant elements of $\gd$.
Moreover, under $\de_L$ they transform according to an irreducible 
representation $\la$ and at least one of them is nonzero (the condition 4.).
 Thus they are linearly independent. They generate (see the
condition 5.) the left module $\gd$. Using Theorem 2.3 of \cite{W2}, we get
that $\dot{p_s}$ ($s\in\spi$) form a linear basis of $\gd_{inv}$ and thus a
basis of the left module $\gd$. Moreover, the same Theorem implies 
that\footnote{there is a missprint in (2.33) of \cite{W2},
to get the correct form one has to replace
$f_{ij}$ by $f_{ij}\circ\kappa^{-2}$, which is denoted by $f_{ij}$
in the present paper (cf \cite{PW})}
\be \dot{p_s}a=(a*f_{st})\dot{p_t} \e{2.1}\ee
for some functionals $f_{st}\in\api'$ such that 
\be f_{st}(ab)=f_{sm}(a)f_{mt}(b),\quad a,b\in\api,\quad
f_{st}(I)=\del_{st}.\e{2.2}\ee 
(It implies
\be f_{ab}\circ S*f_{bc}=f_{ab}*f_{bc}\circ S=\del_{ac}\eps\e{2.2a}\ee
-- one can apply both sides to $v_{ij}$, $v\in\Rep\ \api$, which span $\api$).
Applying $\de_L$ to \r{2.1}, we get
\[ \la_{st}a^{(1)}\ot(a^{(2)}*f_{tr})\dot{p_r}=(\la_{st}\ot\dot{p_t})\de a= \]
\[ \de(a*f_{st})(\la_{tr}\ot\dot{p_r})=(a^{(1)}*f_{st})\la_{tr}\ot
a^{(2)}\dot{p_r}, \]
where we denoted $\de a=a^{(1)}\ot a^{(2)}$. Comparing the coefficients
multiplying $\dot{p_r}$ and applying $\id\ot\eps$, one obtains 
\be \la_{st}(f_{tr}*a)=(a*f_{st})\la_{tr},\quad a\in\api.\e{2.3}\ee

Let us pass to $\g$. The elements $\{I, p_s:\quad s\in\spi\}$ form a basis of
$\g$ as a left module. Moreover, \r{2.1} implies
\be p_s a=(a*f_{st})p_t+\phi_s(a),\quad a\in\api,\e{2.4}\ee
for some $\phi_s:\api\lra\api$. Using $p_s(ab)=(p_sa)b$, $p_sI=p_s$ and
\r{2.2}, one gets 
\be \phi_s(ab)=(a*f_{st})\phi_t(b)+\phi_s(a)b,\quad a,b\in\api,\quad
\phi_s(I)=0.\e{2.5}\ee 
Thus the mapping 
\be \psi:\api\ni a\lra\left(\ba{cc} a*f & \phi(a)\\ 0 & a\ea\right)\in
M_{|\spi|+1}(\api)\e{2.6}\ee 
is a unital homomorphism. Applying $\de$ to \r{2.4}, one gets the equality of
both sides if and only if \r{2.3} and
\be \de\phi_s(a)=(\la_{st}\ot I)[(\id\ot\phi_t)\de(a)]+(\phi_s\ot\id)\de
a,\quad a\in\api,\e{2.7}\ee
hold. 

Before investigation (for certain class of $\api$) of this equation let 
us consider the general situation.
We assume that $f_{st}\in\api'$ and $\phi_s:\api\lra\api$
satisfy \r{2.2}, \r{2.3}, \r{2.5} and \r{2.7}. 
Let $\tb$ be the algebra with $I$ generated by $\api$ and $\te p_s$,
$s\in\spi$, with relations \r{2.4}. We set 
$\te p_K=\te p_{k_1}\cdot\k\cdot\te p_{k_n}$ for $K=(k_1,\k,k_n)\in
\spi_n=\spi\times\k\times \spi$ ($n$ times), 
$\hat \spi=\sqcup^{\infty}_{n=0} \spi_n$ ($\te
p_{\emptyset}=I$). 

\bl\e{l2.1} $\te p_K$, $K\in\hat \spi$, form a basis of the left $\api$-module 
$\tb$. \el

\bd We define $\cpi$ as a free left $\api$-module with basis 
$P_K=p_{k_1}\ot\k\ot p_{k_n}$ for $K=(k_1,\k,k_n)\in\hat\spi$. We also set
$\cpi_n=\api\cdot\spa\{P_K:\ \ K\in \spi_n\}$ and introduce linear mappings
$\lam_n:\cpi_n\ot\api\lra\cpi_n$ by $\lam_0(bP_{\emptyset}\ot a)=
baP_{\emptyset}$,
\[ \lam_n(b(P_K\ot p_i)\ot a)=\lam_{n-1}(bP_K\ot(a*f_{is}))\ot
p_s+\lam_{n-1}(bP_K\ot\phi_i(a)), \]
$K\in \spi_{n-1}$. Next we define linear mapping (multiplication) 
$m:\cpi\ot\cpi\lra\cpi$ in $\cpi$ by 
$m(bP_K\ot aP_L)=\lam_n(bP_K\ot a)\ot P_L$ for
$K,L\in\hat\spi$. After some calculations (using \r{2.5}) one can check that
$(\cpi,m)$ is an algebra with identity $P_{\emptyset}$. Moreover, there 
exists a unital homomorphism
$\rho:\tb\lra\cpi$ given by $\rho(a\te p_K)=aP_K$ (\r{2.4}
holds in $\cpi$). But $P_K$ form a basis of $\cpi$ and hence $\te p_K$ are
independent (over $\api$) in $\tb$. They also generate (due to \r{2.4})
$\tb$ as the left module.\ed

Let the comultiplication in $\tb$ be given by the comultiplication in 
$\api$ and \r{2.0} (it is well defined due to \r{2.3} and \r{2.7} -- see 
remarks before \r{2.7}). Then $\tb$ is a bialgebra with a natural bialgebra
epimorphism $\pi:\tb\lra\bpi$ given by $\pi(a)=a$, $a\in\api$, 
$\pi(\te p_s)=p_s$. 

We set $J=\ker\pi$, $\te\g=\api\cdot\spa\{\te p_s,\ \ s\in\spi\}$, 
$\tb_k=\api\cdot\spa\{\te p_J,\ \ J\in\spi_k\}$ 
(with basis $\te p_J$, $J\in \spi_k$),
$\tb^k=\op^k_{l=0} \tb_l$
(cf. Lemma \re{l2.1}), $J^k=\tb^k\cap J$. Let $J_k$ be some vector
space such that $J^{k-1}\op J_k=J^k$, $k\in\nb$ ($J_0=\{0\}$). We have
$J_k\cap\tb^{k-1}\subset J_k\cap J^{k-1}=\{0\}$, so we can define $\tb_{(k)}$
as some vector space such that $\tb^{k-1}\op J_k\op\tb_{(k)}=\tb^k$. In
particular, we can put $J^0=J^1=J_0=J_1=\{0\}$, 
$J_2=J^2$, $\tb_{(0)}=\tb_{0}=\api$,
$\tb_{(1)}=\tb_1=\te\g$. 

We shall investigate $J_2$. Let $s\in J_2$. Then $s\in J$, 
$(\pi\ot\pi)\de s=\de\pi s=0$, $\de s\in\tb\ot J+J\ot\tb$. But also
$s\in\tb^2$, $\de s\in\tb^2\ot\api+ \api\ot\tb^2+ \tb^1\ot\tb^1$. Using
$\tb=(\op J_k)\op(\op\tb_{(k)})$, $J=\op J_k$,
$\tb^2=\tb_{(0)}\op\tb_{(1)}\op\tb_{(2)}\op J_2$,
$\tb^1=\tb_{(0)}\op\tb_{(1)}$, one gets
\be \de s\in\api\ot J_2\op J_2\ot\api.\e{2.8} \ee

We put $\gd_2=\tb^2/\tb^1$. We see that $\de$ induces a linear mapping
\[ \de_2:\gd_2\lra(\tb^2\ot\api+\api\ot\tb^2+\tb^1\ot\tb^1)/(\tb^1\ot\api+
\api\ot\tb^1)\approx \]
\[ \gd_2\ot\api\op\api\ot\gd_2\op\gd\ot\gd. \]
We get the decomposition $\de_2=\de_{2L}+\de_{2R}+\te\de$, 
$\de_{2L}:\gd_2\lra\api\ot\gd_2$, $\de_{2R}:\gd_2\lra\gd_2\ot\api$ and
$\te\de:\gd_2\lra\gd\ot\gd$. 

\bl\e{l2.2} The elements $[\te p_i\te p_j]$ form a linear basis of
$(\gd_2)_{inv}$, hence a basis of the left module $\gd_2$, while 
$\dot p_i\ot\dot p_j$ form a linear basis of $(\gd\ot_{\api}\gd)_{inv}$, 
hence a basis of the left module $\gd\ot_{\api}\gd$.  

Moreover, 
$\xi:\gd_2\lra\gd\ot_{\api}\gd$ given by
$\xi([\te p_i\te p_j])=\dot{p_i}\ot\dot{p_j}$
defines an isomorphism of bicovariant bimodules.\el

\bd The elements $[\te p_i\te p_j]$ are basis of the left module 
$\gd_2=\tb^2/\tb^1$
(Lemma \re{l2.1}) and belong to $(\gd_2)_{inv}$ 
while $\dot{p_i}\ot\dot{p_j}$ are basis of the left module
$\gd\ot_{\api}\gd$ (they are linearly independent elements of 
$(\gd\ot_{\api}\gd)_{inv}$
and generate the left module). Moreover, 
\[ [\te p_i\te p_j]a=(a*f_{js}*f_{im})[\te p_m\te p_s], \]
\[ \de_{2L}[\te p_i\te p_j]=\la_{im}\la_{js}\ot[\te p_m\te p_s], \]
\[ \de_{2R}[\te p_i\te p_j]=[\te p_i\te p_j]\ot I, \]
and similarly for $\dot{p_i}\ot\dot{p_j}$. Thus 
$(\gd_2,\de_{2L},\de_{2R})$ is a
bicovariant bimodule isomorphic (by $\xi$) to $\gd\ot_{\api}\gd$.\ed

In the following we shall identify $x$ with $\xi(x)$ and $\gd_2$ with
$\gd\ot_{\api}\gd$. Let us recall that there exists a unique bicovariant 
bimodule
isomorphism $\rho:\gd\ot_{\api}\gd\lra\gd\ot_{\api}\gd$ given by
$\rho(\eta\ot\omega)=\omega\ot\eta$ where $\eta$ is a right--invariant, while
$\omega$ is a left--invariant element of $\gd$ ($\rho=\sigma^{-1}$ where
$\sigma$ is given in Proposition 3.1 of \cite{W2}). Thus $\ker(\rho+\id)$ is
 a bicovariant subbimodule of $\gd\ot_{\api}\gd$. 

We define \be R_{ij,sm}=f_{im}(\la_{js}).\e{2.17'}\ee
Setting $a=\la_{mn}$ in \r{2.3}, we get 
\be R\in\mor(\la\tp\la,\la\tp\la).\e{2.17a}\ee
Due to
\be\e{2.9}\rho(\dot{p_i}\ot\dot{p_j})=
f_{im}(\la_{js})\dot{p_s}\ot\dot{p_m}\ee
(similar proof as for (3.15) of \cite{W2})
$R^T$ is the matrix of $\rho$ for the basis $\dot p_i\ot\dot p_j$ 
($i,j\in\spi$) of $(\gd_2)_{inv}$.

\bp\e{p2.2'} Let $K$ be a right--covariant left submodule of 
$\gd_2$, $N=\dim\ K\in\nb$, and 
\be a^{\al}_{ij}(\dot{p_i}\ot\dot{p_j}), \quad \al=1,\k,N, \mbox{ form a 
linear
basis of } K_{inv}\e{2.9a}\ee
($a^{\al}_{ij}\in\cb$). Then $K$ is a bicovariant bimodule iff there
exists $g=(g_{\al\beta})_{\al,\beta=1}^N\in\Rep\ H$ such that
\be a(\la\tp\la)=g\cdot a\e{2.17g}\ee
(we set $a_{\al,mn}=a^{\al}_{mn}$) and $\omega_{\al\beta}\in\api'$,
$\al,\beta=1,\k,N$, such that
\be a^{\al}_{ij}(f_{js}*f_{im})=\omega_{\al\beta}a^{\beta}_{ms},\e{2.29}\ee
\be \omega_{\al\ga}(bc)=\omega_{\al\beta}(b)\omega_{\beta\ga}(c)\quad
(b,c\in\api),\quad \omega_{\al\beta}(I)=\del_{\al\beta}. \e{2.29a}\ee 
In that case $g$ is a quotient representation of $\la\tp\la$, corresponding
to $K_{inv}$:
\be \de_{2L}[a^{\al}_{ij}\te p_i\te p_j]=
g_{\al\beta}\ot[a^{\beta}_{mn}\te p_m\te p_n]. \e{2.217} \ee
Moreover, $K\subset\ker(\rho+\id)$ iff 
\be a^{\al}(R+\jb^{\ot 2})=0,\e{2.18}\ee
where $(a^{\al})_{mn}=a^{\al}_{mn}$.\ep

\bd If $K$ is a bicovariant bimodule then 
$\de_{2L}K_{inv}\subset\api\ot K_{inv}$. Therefore there exist 
$g_{\al\beta}\in\api$ such that \r{2.217} holds. Using the definition and 
properties of $\de_{2L}$, one gets \r{2.17g} and that $g$ is a representation
of $H$. Conversely, \r{2.17g} gives \r{2.217} and left invariance of $K$. 
Moreover, the 
right module condition for $K$ means that for any
$b\in\api$ 
\[ a^{\al}_{ij}[\te p_i\te p_j]b=
a^{\al}_{ij}(b*f_{js}*f_{im})[\te p_m\te p_s]= \]
\[ b_{\al\beta}a^{\beta}_{ms}[\te p_m\te p_s] \]
for some $b_{\al\beta}\in\api$. Setting $\omega_{\al\beta}(b)=e(b_{\al\beta})$ 
we
get $\omega_{\al\beta}\in\api'$ and \r{2.29}--\r{2.29a}  
(we use $(bc)_{\al\ga}=b_{\al\beta}c_{\beta\ga}$, $I_{\al\beta}
=\del_{\al\beta}I$).
Conversely, \r{2.29}--\r{2.29a} give the right module condition for $K$.

Due to \r{2.17g} g is a quotient representation (see Appendix B of \cite{KP})
of $\la\tp\la$ (due to \r{2.9a}, $a$ is surjective). 

Finally, \r{2.9} and \r{2.17'} give 
\[ (\rho+\id)(a^{\al}_{ij}\dot p_i\ot\dot p_j)=
a^{\al}_{ij}(R_{ij,sm}+\del_{is}\del_{jm})\dot p_s\ot\dot p_m \]
and the last statement holds.\ed

{}From now on we set $K=J_2/\tb^1\subset\gd_2$. As left modules $K\approx J_2$ 
since $J_2\cap\tb^1=\{0\}$. We shall see that $K$ satisfies all the
conditions of Proposition \re{p2.2'}:

\bp\e{p2.3} $K$ is a bicovariant subbimodule of
$\ker\te\de=\ker(\rho+\id)\subset\gd\ot_{\api}\gd=\gd_2$.\ep 

\br Thus in interesting situations $\rho$ should have an eigenvalue $-1$
(cf \cite{PW}, \cite{M}).\er

\bd Since $J$ is an ideal, $J_2$ is a bimodule, so is $K$. Due to \r{2.8},
$\de_{2L}K\subset\api\ot K$, $\de_{2R}K\subset K\ot\api$, $\te\de K=0$. 
Therefore
$K\subset\ker\te\de$ is a bicovariant subbimodule of $\gd\ot_{\api}\gd$ (see 
Lemma
\re{l2.2}). It remains to prove $\ker \te\de=\ker(\rho+\id)$. Let
$x=a_{ij}\dot{p_i}\ot\dot{p_j}$. If $x\in \ker\te\de$ then 
\[ 0=\te\de a_{ij}\dot{p_i}\ot\dot{p_j}=\de(a_{ij})[(\dot{p_i}\ot
I)(\la_{js}\ot\dot{p_s})+(\la_{is}\ot\dot{p_s})(\dot{p_j}\ot
I)]_{|_{\gd\ot\gd}}= \]
\[ \de(a_{ij})\{[(\la_{js}*f_{im})+\la_{is}\del_{jm}]\ot
I\}\dot{p_m}\ot\dot{p_s}. \]
Using the independence of $\dot{p_i}$ and acting by $\eps\ot\id$, one gets 
\be\e{2.8a} a_{ij}[f_{im}(\la_{js})+\del_{is}\del_{jm}]=0.\ee
Multiplying from the right by $\dot{p_s}\ot\dot{p_m}$ and using 
\r{2.9}, we obtain 
$(\rho+\id)(x)=(\rho+\id)(a_{ij}\dot{p_i}\ot\dot{p_j})=0$, i.e.
$x\in\ker(\rho+\id)$. Conversely, the last equality implies \r{2.8a}. Acting
by $\de$ and multiplying from the right by $(\la_{sn}\ot
I)(\dot{p_m}\ot\dot{p_n})$, we can get back $\te\de
a_{ij}\dot{p_i}\ot\dot{p_j}=0$, $x\in\ker\te\de$.\ed

We know that \r{2.9a} holds for some $a^{\al}_{ij}\in\cb$. Then 
$a^{\al}_{ij}(\dot{p_i}\ot\dot{p_j})$ ($\al=1,\k,N$) form a basis of
the left module $K$. Let 
\be s^{\al}=a^{\al}_{ij}\te p_i\te p_j+b^{\al}_i\te p_i+c^{\al}\e{2.10}\ee
be the corresponding basis elements of the left module $J_2$
($J_2\cap\tb^1=\{0\}$). We get 

\bp\e{p2.4} As left module $\api\cdot\spa\{p_ip_j,p_i,I:
i,j\in\spi\}\approx\tb^2/J_2$ has dimension $|\spi|^2+|\spi|+1-\dim K$.\ep

\sectio{Properties of inhomogeneous quantum groups}

Here we continue the investigations of the previous Section (assuming the
conditions given at its beginning)  and find the form of commutation relations
 in $\bpi$. 
As before $\po(G)=(\bpi,\de)$ and $\po(H)=(\api,\de)$. Moreover, we assume
\btr
\item[a.] Each representation of $H$ is completely reducible
\item[b.] $\la$ is an irreducible representation of $H$
\item[c.] $\mor(v\tp w,\la\tp v\tp w)=\{0\}$ for any two irreducible
representations $v,w$ of $H$.
\etr
The condition c. is used only for simplicity and will be removed in Remark
\re{r2.99}. 

We return to the investigation of \r{2.7}. Due to the condition a.,
$a=u_{AB}$, $u\in \Irr\ H$, $A,B=1,\k,\dim\ u$, form a basis of $\api$.
Setting $\phi_s(u_{AB})=\phi_{sA,B}$, \r{2.7} is equivalent to
\[ \de\phi_{sA,B}=(\la\tp u)_{sA,tC}\ot\phi_{tC,B}+\phi_{sA,C}\ot u_{CB}. \]
Multiplying both sides from the right by 
$\de(u^c_{DB})=u^c_{DL}\ot u^c_{LB}$ 
(where $u^c_{DB}=u^{-1}_{BD}$ etc.) and setting
$\rho_{sAD}=\phi_{sA,B}u^{-1}_{BD}$, one gets 
\be \de\rho_{sAD}=(\la\tp u\tp
u^c)_{sAD,tCL}\ot\rho_{tCL}+\rho_{sAD}\ot I.\e{2.11}\ee
Therefore
$\left(\ba{cc} \la\tp u\tp u^c & \rho \\ 0 & I \ea\right)$ is a 
representation
of $H$. Using the condition a., there must exist a vector 
$\left(\ba{c}w\\1\ea\right)$corresponding to the representation $I$. It 
means
\be \rho=w-(\la\tp u\tp u^c)w \e{2.12}\ee
(conversely, \r{2.12} implies \r{2.11}). We define $\eta_i\in\api'$ by
$\eta_i(u_{AB})=w_{iAB}$. Using \r{2.12}, we get 
\[ \phi_s(u_{AB})=\phi_{sA,B}=\rho_{sAD}u_{DB}=\eta_s(u_{AD})u_{DB}-(\la\tp
u)_{sA,tL} \eta_t(u_{LB}) \] 
and \r{2.7} is equivalent to 
\be \phi_s(a)=a*\eta_s-\la_{st}(\eta_t*a),\quad a\in\api.\e{2.13}\ee
Due to c. ($v=u$, $w=I$), $\eta_s$ are uniquely determined by $\phi_s$.
Inserting \r{2.13} to \r{2.5}, we obtain $\eta_s(I)=0$ and
\[ ab*\eta_s-\la_{sm}(\eta_m*ab)=\]
\[ (a*f_{st})(b*\eta_t)-
(a*f_{st})\la_{tm}(\eta_m*b)+(a*\eta_s)b-\la_{st}(\eta_t*a)b, \]
which (see \r{2.3}) we can write as 
\be\left.\ba{l} ab*\eta_s-(a*f_{st})(b*\eta_t)-(a*\eta_s)b=\\\\
\la_{sm}[\eta_m*ab-(f_{mt}*a)(\eta_t*b)-(\eta_m*a)b].\ea\right\}\e{2.14}\ee
Setting $L^{vw}_{mAB,CD}=\eta_m(ab)-f_{mt}(a)\eta_t(b)-\eta_m(a)\eps(b)$ for
$a=v_{AC}$, $b=w_{BD}$, $v,w\in \Irr\ H$, we can replace \r{2.14} by 
$L^{vw}\in\mor(v\tp w,\la\tp v\tp w)=\{0\}$, so \r{2.7} is equivalent to
\be \eta_m(ab)=\eta_m(a)\eps(b)+f_{mt}(a)\eta_t(b),\quad a,b\in\api,\quad
\eta_m(I)=0.\e{2.15}\ee
Combining \r{2.2} with \r{2.15},
\be \api\ni a\lra\rho(a)=
\left(\ba{cc} f(a) & \eta(a)\\ 0 & \eps(a)\ea\right)\in
M_{|\spi|}(\cb) \mbox{ is a unital homomorphism. }\e{2.16}\ee
 We get

\bt\e{t2.5} Let $\api$ be a Hopf algebra satisfying a.--c.. Then the general
bialgebra $\bpi$ satisfying the conditions 1.-5. is equal $\tb/J$ where 
$\tb$
is the algebra with $I$ generated by $\api$ and $\te p_s$ ($s\in\spi$) with
relations \r{2.4} where $\phi_s$ is given by \r{2.13} for $f$ and $\eta$
satisfying \r{2.3} and \r{2.16}. Moreover, $\tb$ is a bialgebra with
comultiplication given by the comultiplication in $\api$ and \r{2.0}. $J$ is
an ideal in $\tb$ such that $\de J\subset J\ot\tb+\tb\ot J$, $\eps(J)=0$,
$J\cap\tb^1=\{0\}$. Conversely, each such $f$, $\eta$ and $J$ give 
$\bpi=\tb/J$
satisfying the conditions 1.-5..\et

\bd It follows from the previous considerations.\ed

Let us recall that $s^{\al}$, $\al=1,\k,N=\dim\ K$, form a basis of the left 
module
$J_2=J\cap\tb^2$. Due to \r{2.10} and \r{2.8}, 
\be\left.\ba{c} \de s^{\al}=a^{\al}_{ij}\la_{im}\la_{jn}\ot\te p_m\te p_n+
a^{\al}_{ij}\te p_i\te p_j\ot I+\\\\
a^{\al}_{ij}(\la_{im}\ot\te p_m)(\te p_j\ot I)+ 
a^{\al}_{ij}(\te p_i\ot I)(\la_{jn}\ot\te p_n)+\\\\
\de(b^{\al}_i)(\la_{ij}\ot\te p_j)+\de(b^{\al}_i)(\te p_i\ot I)+\de(c^{\al})
\in\\\\ 
\api\ot J_2\op
J_2\ot\api.\ea\right\}\e{2.17}\ee
In particular, the terms in $\te\g\ot\te\g$ should cancel out, which is 
equivalent (cf the proof of Proposition \re{p2.3}) to \r{2.8a} for $a=a^{\al}$,
i.e. to \r{2.18}. The equations \r{2.18} are down to earth formulation of
the condition $K\subset\ker(\rho+\id)$. Using that and \r{2.17g}, one gets
\[ \de s^{\al}-s^{\al}\ot I-
g_{\al\beta}\ot s^{\beta}\in(\api\ot J_2\op J_2\ot\api)
\cap(\api\ot\tb^1+\tb^1\ot\api)=\{0\}. \]
Thus \r{2.17} is equivalent to
\[ g_{\al\beta}\ot[-b^{\beta}_i\te p_i-c^{\beta}]+
[-b^{\al}_i\te p_i-c^{\al}]\ot I+\]
\[ a^{\al}_{ij}\phi_i(\la_{jn})\ot\te p_n+ 
\de(b^{\al}_i)(\la_{ij}\ot\te p_j)+
\]
\[ \de(b^{\al}_i)(\te p_i\ot I)+ \de(c^{\al})=0.\]
Using Lemma \re{l2.1},
\be \de(b^{\al}_i)=b^{\al}_i\ot I,\e{2.19}\ee
\be \de(b^{\al}_i)\{[\la_{ij}-g_{\al\beta}b^{\beta}_j+
a^{\al}_{ik}\phi_i(\la_{kj})]\ot I\}=0,\e{2.20}\ee
\be \de(c^{\al})=g_{\al\beta}\ot c^{\beta}+c^{\al}\ot I.\e{2.21}\ee
In virtue of \r{2.19}, $b^{\al}_i\in\cb$. Using \r{2.13}
and \r{2.17g}, we can write
\r{2.20} as
\[ [b^{\al}_s+a^{\al}_{ik}\eta_i(\la_{ks})]\la_{sj}=
g_{\al\beta}[b^{\beta}_j+a^{\beta}_{mr}\eta_m(\la_{rj})]. \]
Using the condition c. for $v=\la$, $w=I$ (according to the condition a., $g$
is equivalent to a subrepresentation of $\la\tp\la$), we get
$\mor(\la,g)=\{0\}$ and hence \r{2.20} is equivalent to
\be b^{\al}_s=-a^{\al}_{ik}\eta_i(\la_{ks}).\e{2.22}\ee

Decomposing $g$ into a direct sum of irreducible representations, 
$c^{\al}$ has 
also
a decomposition into a direct sum. So we can solve \r{2.21}  in each
irreducible component and thus assume that $g$ is irreducible. For any 
$v\in\ \Irr\ H$ we set $\api_v=\spa\{v_{mn}:m,n=1,\k,\dim\ v\}$. We know
$\api=\op_{v\in \Irr\ H}\api_v$.\\
i) $g=I$, hence $\de(c^{\al})=I\ot c^{\al}+c^{\al}\ot I$. Thus 
\[ \de c^{\al}\in(\op_{v\in \Irr\ H} \api_v\ot\api_v)\cap
(\op_{v\in \Irr\ H} (\api_I\ot\api_v\op\api_v\ot\api_I))=\]
\[ \api_I\ot\api_I=\cb I\ot I,\]
$c^{\al}=\lam I$, $\lam\in\cb$. That gives $\lam=2\lam$, $\lam=0$, 
$c^{\al}=0$.\\
ii) $g\in \Irr\ H$, $g\not\simeq I$. Then $c^{\al}=\sum c^{\al}_v$,
$c^{\al}_v\in\api_v$ and \r{2.21} is equivalent to 
\[ \de c^{\al}_I=c^{\al}_I\ot I\in\api_I\ot\api_I,\]
\[ 0=g_{\al\beta}\ot c^{\beta}_I+
c^{\al}_g\ot I\in\api_g\ot\api_I,\]
\[ \de c^{\al}_g=g_{\al\beta}\ot c^{\beta}_g\in\api_g\ot\api_g,\]
\[ \de c^{\al}_v=0\in\api_v\ot\api_v,\qquad
0=g_{\al\beta}\ot c^{\beta}_v\in\api_g\ot\api_v, \]
\[ 0=c^{\al}_v\ot I\in\api_v\ot\api_I,\quad v\in\ \Irr\ H,\quad v\not\simeq
I,g.\]
Solving these relations, one gets $c^{\al}_I\in\cb$,
$c^{\al}_g=-g_{\al\beta}c^{\beta}_I$, $c^{\al}_v=0$ for $v\not\simeq I,g$,
$v\in \Irr\ H$. Then $c^{\al}=c^{\al}_I-g_{\al\beta}c^{\beta}_I$. It holds 
also in the case i) and for whole $g$. 

Since $a^{\al}$ are linearly independent, there exist $T_{mn}\in\cb$ such 
that
$c^{\al}_I=a^{\al}_{mn}T_{mn}$. One gets $c^{\al}=a^{\al}_{mn}
(T_{mn}-\la_{ma}\la_{nb}T_{ab})$ (we have used \r{2.17g}). Concluding,
\be s^{\al}=a^{\al}_{ij}(\te p_i\te p_j-\eta_i(\la_{js})\te p_s+T_{ij}-
\la_{im}\la_{jn}T_{mn})\e{2.26}\ee
and we get ($N=\dim\ K$)

\bt\e{t2.6} Let $\bpi$ be as in Theorem \re{t2.5}. Then $J_2=J\cap\tb^2$ is an
$\api$-bimodule
and as the left module it has a basis \r{2.26}, $\al=1,\k,N$, for some 
$a^{\al}_{ij},\ T_{ij}\in\cb, N\in\nb$. 
Moreover, $a^{\al}$ satisfy \r{2.18}, \r{2.17g} and \r{2.29}--\r{2.29a}
for some $g\in\Rep\ H$ and $\omega_{\al\beta}\in\api'$ ($\al,\beta=1,\k,N$).\et

\bt\e{t2.7} Let $j_2\subset\tb^2$ be the left module generated by \r{2.26} 
for 
some $a^{\al}_{ij}$, $T_{ij}\in\cb$, such that $a^{\al}$ 
($\al=1,\k,N$) are linearly
independent and satisfy \r{2.18}, \r{2.17g} and \r{2.29}--\r{2.29a}
for some $g\in\Rep\ H$ and $\omega_{\al\beta}\in\api'$. Then 
\be \de j_2\subset (j_2\ot\api)\op(\api\ot j_2),\qquad
j_2\cap\tb^1=\{0\}.\e{2.27}\ee
Moreover, $j_2$ is a bimodule if and only if 
\be g_{\al\beta}(\tau_{\beta}*b)=b*\tau_{\al},\quad b\in\api,\e{2.28}\ee
where $\tau_{\al}=a^{\al}_{ij}\tau_{ij},$
\[ \tau_{ij}=
\eta_j*\eta_i-\eta_i(\la_{js})\eta_s+T_{ij}\eps-(f_{jn}*f_{im})T_{mn} \]
and $g$ is given by \r{2.17g}.\et

\bd The first statement follows from the computations before Theorem 
\re{t2.6}.
Due to Proposition \re{p2.2'}, $j_2/\tb^1$ is a bicovariant bimodule
contained in $\ker(\rho+\id)$. In order to prove the last statement 
we compute
\be\left.\ba{l} s^{\al}b=
a^{\al}_{ij}[\te p_i\te p_j-
\eta_i(\la_{js})\te p_s+T_{ij}-\la_{im}\la_{jn}T_{mn}]b= \\\\
 a^{\al}_{ij}\{\te p_i[(b*f_{js})\te p_s+\phi_j(b)]-\eta_i(\la_{js})\te p_sb+
(T_{ij}-\la_{im}\la_{jn}T_{mn})b\}= \\\\
 a^{\al}_{ij}\{(b*f_{js}*f_{im})\te p_m\te p_s+\phi_i(b*f_{jr})\te p_r+
[\phi_j(b)*f_{ir}]\te p_r+ \\\\
 \phi_i(\phi_j(b))-\eta_i(\la_{js})(b*f_{sr})\te p_r-
\eta_i(\la_{js})\phi_s(b)+\\\\ 
 (T_{ij}-\la_{im}\la_{jn}T_{mn})b\}.\ea\right\}\e{2.30} \ee
Using \r{2.29}, we get
\be (b*\xi_{\al\beta})s_{\beta}=a^{\al}_{ij}(b*f_{jn}*f_{im})
(\te p_m\te p_n-\eta_m(\la_{nr})\te p_r+T_{mn}-
\la_{ma}\la_{nb}T_{ab}),\e{2.30a}\ee
hence
\be s^{\al}b-(b*\xi_{\al\beta})s^{\beta}=A_{\al r}\te p_r+B_{\al}\e{2.31}\ee
for some $A_{\al r}$, $B_{\al}\in\api$. We conclude that $j_2$ is a right
module if and only if  \r{2.31} belongs to $j_2$ for any $b$, which means
$A_{\al r}=B_{\al}=0$ ($j_2\cap\tb^1=\{0\}$).

Using \r{2.30}, \r{2.30a}, \r{2.31}, \r{2.13} and \r{2.2}, one obtains
\[ A_{\al r}=a^{\al}_{ij}\{b*f_{jr}*\eta_i-\la_{im}(\eta_m*b*f_{jr})+ \]
\[ b*\eta_j*f_{ir}-(\la_{js}*f_{im})(\eta_s*b*f_{mr})-\]
\[ \eta_i(\la_{js})(b*f_{sr})+ (b*f_{jn}*f_{im})\eta_m(\la_{nr})\}. \]
In virtue of \r{2.18}
\[ a^{\al}_{ij}(\la_{js}*f_{im})=a^{\al}_{ij}f_{im}(\la_{jk})\la_{ks}= \] 
\[ a^{\al}_{ij}R_{ij,km}\la_{ks}=-a^{\al}_{km}\la_{ks} \]
so the second and the fourth terms in $A_{\al r}$ cancel. On the other hand, 
\r{2.3}, \r{2.15}  imply 
\be\left.\ba{c} 0=\eta_i\{(b*f_{js})\la_{sr}-\la_{js}(f_{sr}*b)\}= \\\\
 \eta_i(b*f_{jr})+f_{im}(b*f_{js})\eta_m(\la_{sr})- \\\\
 \eta_i(\la_{js})f_{sr}(b)-
f_{im}(\la_{js})\eta_m(f_{sr}*b),\ea\right\}\e{2.31a}\ee
hence due to  \r{2.18}
\be\left.\ba{c} a^{\al}_{ij}[(f_{jr}*\eta_i)(b)+
(f_{js}*f_{im})(b)\eta_m(\la_{sr})- \\\\
 \eta_i(\la_{js})f_{sr}(b)+(\eta_j*f_{ir})(b)]=0.\ea\right\}\e{2.31a'}\ee
Therefore also other terms in $A_{\al r}$ vanish, $A_{\al r}=0$ for each
$b\in\api$. 

In virtue of \r{2.13}, \r{2.5} and \r{2.3},
\[ B_{\al}=a^{\al}_{ij}[\phi_i(b*\eta_j)-(\la_{jm}*f_{is})\phi_s(\eta_m*b)- 
\]
\[ \phi_i(\la_{js})(\eta_s*b)-\eta_i(\la_{js})\phi_s(b)+
(T_{ij}-\la_{im}\la_{jn}T_{mn})b- \]
\[ (b*f_{jn}*f_{im})(T_{mn}-\la_{ma}\la_{nb}T_{ab})]= \]
\[ a^{\al}_{ij}\cdot[b*\eta_j*\eta_i-\la_{im}(\eta_m*b*\eta_j)- \]
\[ f_{is}(\la_{jl})\la_{lm}(\eta_m*b*\eta_s)+
f_{is}(\la_{jl})\la_{lm}\la_{sn}(\eta_n*\eta_m*b)- \]
\[ \eta_i(\la_{jm})\la_{ms}(\eta_s*b)+
\la_{im}\la_{jn}\eta_m(\la_{ns})(\eta_s*b)- \]
\[ \eta_i(\la_{js})(b*\eta_s)+\eta_i(\la_{js})\la_{sm}(\eta_m*b)+ \]
\[ T_{ij}b-\la_{im}\la_{jn}T_{mn}b-(b*f_{jn}*f_{im})T_{mn}+ \]
\[ \la_{im}\la_{jn}(f_{nc}*f_{ma}*b)T_{ac}]. \]
The fifth and the eight terms cancel. Using \r{2.18}, the second and the 
third
terms also give 0. The terms 1,7,9 and 11 produce $b*\tau_{\al}$. In virtue 
of
$a^{\al}_{ij}\la_{im}\la_{jn}=g_{\al\beta}a^{\beta}_{mn}$ (see \r{2.17g}) and
\r{2.18} the terms 4,6,10 and 12 yield $-g_{\al\beta}(\tau_{\beta}*b)$. Thus
$B_{\al}=b*\tau_{\al}-g_{\al\beta}(\tau_{\beta}*b)$ and our Theorem 
follows.\ed

Using the notation of Theorem \re{t2.7} one has

\bp\e{p2.7'} 
\[ \tau_{\al}(ab)=\omega_{\al\beta}(a)\tau_{\beta}(b)+\tau_{\al}(a)\eps(b)\quad
(a,b\in\api),\quad \tau_{\al}(I)=0. \]\ep

\bd We have (see \r{2.2}, \r{2.15}) 
\[ \tau_{ij}(ab)=(\eta_j*\eta_i)(ab)-
\eta_i(\la_{js})[f_{sr}(a)\eta_r(b)+\eta_s(a)\eps(b)]+ \]
\[ T_{ij}\eps(a)\eps(b)-(f_{js}*f_{im})(a)(f_{sr}*f_{ml})(b)T_{lr}. \]
But (we use \r{2.15},  \r{2.31a'} and \r{2.29})
\[ a^{\al}_{ij}(\eta_j*\eta_i)(ab)=
a^{\al}_{ij}[(\eta_j*\eta_i)(a)\eps(b)+(f_{jr}*\eta_i)(a)\eta_r(b)+
\]
\[ (\eta_j*f_{ir})(a)\eta_r(b)+(f_{jr}*f_{is})(a)(\eta_r*\eta_s)(b)]= \]
\[ a^{\al}_{ij}[(\eta_j*\eta_i)(a)\eps(b)+\eta_i(\la_{js})f_{sr}(a)\eta_r(b)- 
\]
\[ (f_{js}*f_{im})(a)\eta_m(\la_{sr})\eta_r(b)+
(f_{js}*f_{im})(a)(\eta_s*\eta_m)(b)]= \]
\[ a^{\al}_{ij}[(\eta_j*\eta_i)(a)\eps(b) 
+ \eta_i(\la_{js})f_{sr}(a)\eta_r(b)]+ \]
\[ \omega_{\al\beta}(a)a^{\beta}_{ms}[(\eta_s*\eta_m)(b)-
\eta_m(\la_{sr})\eta_r(b)]. \]
Combining these facts, we get the first assertion.
The second one is trivial.\ed

\br Proposition \re{p2.7'} and \r{2.29a} give that
\[ \zeta:\api\ni a\lra
\left(\ba{cc} \omega(a) & \tau(a)\\ 0 & \eps(a)\ea\right)\in M_{N+1}(\cb) \]
is a unital homomorphism, where
$\omega(a)=(\omega_{\al\beta}(a))_{\al,\beta=1}^N$,
$\tau(a)=(\tau_{\al}(a))_{\al=1}^N$.\er

\br\e{r2.220}
Let $S$ be a set generating $\api$ as algebra with unity. One can prove that
\r{2.31a'} for $b\in S$ implies \r{2.31a'} for $b\in\api$ (due to
\r{2.31} $A_{\al r}=0$ for $b,b'$ implies $A_{\al r}=0$ for $bb'$).
Similarly, \r{2.28} for $b\in S$ implies \r{2.28} for $b\in\api$
(it is equivalent to the right module condition which suffices to check 
only for $b\in S$).\er

\sectio{Structure of inhomogeneous quantum groups}

Here  we 
continue the investigations of two preceding Sections 
(including the assumptions made at their beginnings) and
find the exact form and `size' of inhomogenous quantum groups.
{}From now on 
we shall consider the most natural situation (which is the case for \qpg s):
\[ R^2=\jb^{\ot 2}\quad\mbox{ and }\quad 
x(R+\jb^{\ot 2})=0\iffi x\in\spa\{a^{\al}:\ \al=1,\k,\dim\ K\} \]
(cf \r{2.17'} and \r{2.18}). In other words: $\rho^2=\id$ and
$K=J_2/\tb^1=\ker(\rho+\id)$. The second condition means that we have
as many relations $a^{\al}_{ij}p_ip_j+b^{\al}_ip_i+c^{\al}=0$ as it is 
allowed by \r{2.18} (for $b^{\al}_{i}=c^{\al}=0$ $p_i$ would be R-symmetric:
$R_{kl,ij}p_ip_j=p_kp_l$). 

We set 
\[A_3=\jb\ot\jb\ot\jb-R\ot\jb-\jb\ot R+(R\ot\jb)(\jb\ot R)+ \]
\[ (\jb\ot R)(R\ot\jb)-(R\ot\jb)(\jb\ot R)(R\ot\jb), \]
$F_{ijk,m}=\tau_{ij}(\la_{km})$, 
$Z_{ij,m}=\eta_i(\la_{jm})$. The main result of the Section is contained in

\bt\e{t2.8} Let $f,\eta$ satisfy \r{2.3}, \r{2.16} and $\rho^2=\id$. The
following conditions are equivalent:
\btr
\item[i)] $J$ is as in Theorem \re{t2.5} with $K=J_2/\tb^1=\ker(\rho+\id)$
\item[ii)] $J$ is the ideal generated by
\[ s_{kl}=(R-\jb^{\ot 2})_{kl,ij}(\te p_i\te p_j-\eta_i(\la_{js})\te p_s+
T_{ij}-\la_{im}\la_{jn}T_{mn}) \]
\mbox{}\ \ \ \ \ \ \ \ for some 
complex numbers $\{T_{ij}\}_{i,j\in\spi}$ satisfying \r{2.28},
\be A_3F=0, \e{2.31b} \ee
\be A_3(Z\ot\jb-\jb\ot Z)T\in\mor(I,\la\tp\la\tp\la). \e{2.32} \ee\etr
If the condition i) or ii) is satisfied then
\be J_2=\api\cdot\spa\{s_{kl}:\ k,l\in\spi\} \e{2.33}\ee
and $\bpi=\tb/J$ satisfies the conditions 1.-5..\et

\noindent {\em Proof.} One has 
\be R^2=\jb^{\ot 2}. \e{2.34}\ee
Let $J$ be as in Theorem \re{t2.5}, $J_2=J\cap\tb^2$ and $K=J_2/\tb^1=
\ker(\rho+id)$. All the conditions of Proposition \re{p2.2'} are
satisfied in that case. Theorem \re{t2.6} and Theorem \re{t2.7} give
\r{2.28}. Moreover (cf the beginning of the Section),
\be \spa\{a^{\al}:\ \al=1,\k,\dim K\}=\spa\{a^{kl}:\ k,l=1,\k,|\spi|\},
\e{2.34a}\ee
where $(a^{kl})_{ij}=(R-\jb^{\ot 2})_{kl,ij}$. 
Thus \r{2.33} is satisfied (see Theorem \re{t2.6}).
Hence, in $\bpi$
\[ p_kp_l=R_{kl,ij}p_ip_j+r_{kl}, \]
where 
\be r_{kl}=c_{kl,s}p_s+M_{kl},\e{2.35}\ee
\be c_{kl,s}=-(R-\jb^{\ot 2})_{kl,ij}\eta_i(\la_{js}) \e{2.36}\ee
(i.e. $c=-(R-\jb^{\ot 2})Z$),
\be M_{kl}=(R-\jb^{\ot 2})_{kl,ij}(T_{ij}-W_{ij}), \e{2.37}\ee
\be W_{ij}=\la_{im}\la_{jn}T_{mn}. \e{2.38}\ee
In short,
\be p\tp p=R(p\tp p)+r.\e{2.39}\ee
Therefore
\[ p\tp p\tp p=(p\tp p)\tp p=(R\ot\jb)(p\tp(p\tp p))+r\tp p= \]
\[ (R\ot\jb)[(\jb\ot R)((p\tp p)\tp p)+p\tp r]+r\tp p= \]
\[ (R\ot\jb)(\jb\ot R)(R\ot\jb)(p\tp p\tp p)+(R\ot\jb)(\jb\ot R)(r\tp p)+ \]
\[ (R\ot\jb)(p\tp r)+r\tp p. \]
On the other hand,
\[ p\tp p\tp p=p\tp(p\tp p)=(\jb\ot R)((p\tp p)\tp p)+p\tp r= \]
\[ (\jb\ot R)[(R\ot\jb)(p\tp(p\tp p))+r\tp p]+p\tp r= \]
\[ (\jb\ot R)(R\ot\jb)(\jb\ot R)(p\tp p\tp p)+(\jb\ot R)(R\ot\jb)(p\tp r)+ \]
\[ (\jb\ot R)(r\tp p)+p\tp r. \]
But the braid equation for $\rho$ (see (3.8) of \cite{W2}) implies
\be (R\ot\jb)(\jb\ot R)(R\ot\jb)=(\jb\ot R)(R\ot\jb)(\jb\ot R).\e{2.40}\ee
Thus
\be A(r\tp p)=B(p\tp r),\e{2.41}\ee
where
\be A=(R\ot\jb)(\jb\ot R)-\jb\ot R+\jb\ot\jb\ot\jb,\e{2.42}\ee
\be B=(\jb\ot R)(R\ot\jb)-R\ot\jb+\jb\ot\jb\ot\jb.\e{2.43}\ee
Using $r=cp+M$ and \r{2.4}, we can rewrite \r{2.41} as
\be H(p\tp p)+Lp+N=0,\e{2.44}\ee
where 
\be H=A(c\ot\jb)-B(\jb\ot c),\e{2.44a}\ee
\be L_{ijk,s}=A_{ijk,mns}M_{mn}-B_{ijk,mnl}(M_{nl}*f_{ms}),\e{2.45}\ee
\be N_{ijk}=-B_{ijk,mnl}\phi_m(M_{nl}).\e{2.46}\ee
Therefore (see \r{2.33}), \be H(\te p\tp\te p)+L\te p+N=
D[(R-\jb^{\ot 2})(\te p\tp\te p)+c\te p+M]\e{2.47}\ee
for some matrix $D$, which thus satisfies $H=D(R-\jb^{\ot 2})$. It exists iff
\be H(R+\jb^{\ot 2})=0 \e{2.48}\ee
and can be chosen as $D=-\frac12 H$. Consequently, \r{2.41} is equivalent to
\r{2.48},
\be L=-\frac12 Hc,\e{2.49}\ee
and
\be N=-\frac12 HM.\e{2.50}\ee

Let us now consider \r{2.48}-\r{2.50} as abstract conditions
for $\eta_i$, $T_{kl}$. 
We shall prove that \r{2.48} follows from the previous conditions.
By virtue of \r{2.31a'} for $b=\la_{kl}$,
\[ (R-\jb^{\ot 2})_{nt,ij}[f_{jr}(\la_{ka})\eta_i(\la_{al})+
f_{js}(\la_{ka})f_{im}(\la_{al})\eta_m(\la_{sr})- \]
\[ \eta_i(\la_{js})f_{sr}(\la_{kl})+\eta_j(\la_{ka})f_{ir}(\la_{al})]=0, \]
hence
\be\left.\ba{c} 
[(R-\jb^{\ot 2})\ot\jb]\{(\jb\ot R)(Z\ot\jb)+(\jb\ot R)(R\ot\jb)(\jb\ot Z)-\\\\
(Z\ot\jb)R+(\jb\ot Z)R\}=0.\ea\right\}\e{2.51}\ee
Multiplying from the left by $A$ and using
\be A((\jb^{\ot 2}-R)\ot\jb)=B(\jb\ot(\jb^{\ot 2}-R))=A_3,\e{2.53}\ee
\be A_3(\jb\ot R)=-A_3,\qquad A_3(R\ot\jb)=-A_3\e{2.52}\ee
(it follows from \r{2.34} and \r{2.40}), we get
\be A_3(\jb\ot Z-Z\ot\jb)(\jb^{\ot 2}+R)=0.\e{2.52a}\ee
But in virtue of \r{2.44a} and \r{2.53}
\be H=A_3(Z\ot\jb-\jb\ot Z)\e{2.54}\ee
and \r{2.48} follows.

Now we shall consider \r{2.49}. One has
\[ W_{ij}*f_{ms}=(\la_{ia}\la_{jb}T_{ab})*f_{ms}= \]
\[ f_{mn}(\la_{ic})f_{ns}(\la_{jd})\la_{ca}\la_{db}T_{ab}=
[(R\ot\jb)(\jb\ot R)(W\ot\jb)]_{mij,s}, \]
\[ T_{ij}*f_{ms}=\del_{ms}T_{ij}=(\jb\ot T)_{mij,s}, \]
hence (see \r{2.45}, \r{2.37}, \r{2.53})
\be L=A_3((W-T)\ot\jb)-A_3(W\ot\jb)+A_3(\jb\ot T)=A_3(\jb\ot T-T\ot\jb).
\e{2.54a}\ee
Using \r{2.48} and \r{2.54},
\be \frac12 Hc=HZ=A_3(Z\ot\jb-\jb\ot Z)Z.\e{2.55}\ee
Moreover,
\[ F_{ijk,m}=\tau_{ij}(\la_{km})=\eta_j(\la_{ks})\eta_i(\la_{sm})- \]
\[ \eta_i(\la_{js})\eta_s(\la_{km})+T_{ij}\del_{km}-
f_{jn}(\la_{ks})f_{ir}(\la_{sm})T_{rn}. \]
Thus
\be F=(\jb\ot Z)Z-(Z\ot\jb)Z+T\ot\jb-(\jb\ot R)(R\ot\jb)(\jb\ot T),
\e{2.55a}\ee 
\[ -A_3F=\frac12 Hc+L, \]
and \r{2.49} is equivalent to \r{2.31b}.

Finally, we investigate \r{2.50}.
According to \r{2.5} and \r{2.13}, 
\[ \phi_m(W_{ij})=(\la_{ia}*f_{ms})\phi_s(\la_{jb})T_{ab}+
\phi_m(\la_{ia})\la_{jb}T_{ab}, \]
\[ \phi_s(\la_{jb})=\eta_s(\la_{jc})\la_{cb}-\la_{sr}\la_{jk}\eta_r(\la_{kb}).
\]
Setting $X_{mij}=\phi_m(W_{ij})$, one gets 
\[ X=(R\ot\jb)(\jb\ot Z)W+(Z\ot\jb)W-(R\ot\jb)(\la\tp\la\tp\la)(\jb\ot Z)T-
(\la\tp\la\tp\la)(Z\ot\jb)T. \]
Using \r{2.46}, \r{2.37}, \r{2.53} and \r{2.17a}, we obtain
\[ N=-A_3X=-A_3(Z\ot\jb-\jb\ot Z)W+(\la\tp\la\tp\la)A_3(Z\ot\jb-\jb\ot Z)T. \]
Due to \r{2.48} and \r{2.54}
\[ \frac12 HM=A_3(Z\ot\jb-\jb\ot Z)(W-T). \]
Therefore
\[ N+\frac12 HM=(\la\tp\la\tp\la)m-m, \]
where $m=A_3(Z\ot\jb-\jb\ot Z)T$ and \r{2.50} is equivalent to \r{2.32}.
Thus the condition i) implies 
\r{2.33}, \r{2.28}, \r{2.31b} 
and
\r{2.32} for some complex numbers $T_{ij}$ ($i,j\in\spi$).

Let us now assume \r{2.28}, \r{2.31b} and \r{2.32}. We define $j_2$  
as the right hand side of \r{2.33}.
Thus $j_2/\tb^1=K=\ker(\rho+\id)$. In virtue of Proposition \re{p2.2'} and
Theorem \re{t2.7} we get \r{2.27} and the bimodule property of $j_2$. 
Let $j$ be the ideal generated by $j_2$. Then $\de j\subset j\ot\tb+\tb\ot j$.
Moreover, 
$\de\te p=\te p\ti I+\la\ti\te p$ implies $e(\te p)=0$, hence $e(j_2)=0$, 
$e(j)=0$. The
previous computations show that \r{2.41} holds in $\tb/j_2$. We shall show
$j\cap\tb^1=\{0\}$, $j\cap\tb^2=j_2$.
Therefore $j$ is as in Theorem \re{t2.5} and $\bpi=\tb/j$ satisfies the
conditions 1.-5.. Furthermore, we will prove that if $J$ satisfies the 
condition i) and $J\cap\tb^2=j_2$ then $J=j$.
Thus the proof of
the Theorem will be finished.

We set $R_k=\jb^{\ot(k-1)}\ot R\ot\jb^{\ot(n-k-1)}$ (cf the notation in the
Introduction), $k=1,2,\k,n-1$, $R_{\pi}=R_{i_1}\cdot\k\cdot R_{i_s}$ for a
permutation $\pi\in\Pi_n$ with a minimal decomposition into transpositions
\be \pi=t_{i_1}\k t_{i_s}.\e{2.56}\ee
Due to \r{2.40}, $R_{\pi}$ is well defined. We set 
\[ S_n=\frac{1}{n!}\sum_{\pi\in\Pi_n} R_{\pi}. \]
Moreover, we put 
$r_{nk}=p^{\tu(k-1)}\tp r\tp p^{\tu(n-k-1)}$ (see \r{2.35}),
\be\left.\ba{ll} r_{n\pi}=r_{ni_1}+R_{i_1}r_{ni_2}+R_{i_1}R_{i_2}r_{ni_3}+
\k+ \\\\
R_{i_1}\cdot\k\cdot R_{i_{s-1}}r_{ni_s},\ea\right\}\e{2.57}\ee
(we choose some decomposition \r{2.56} for each $\pi$), 
$r_n=\frac{1}{n!}\sum_{\pi} r_{n\pi}$. We shall prove the following

\bp\e{p2.8}  Let $j$ be the ideal generated 
by
\r{2.33}. 
We assume \r{2.41} in $\tb/j_2$ and \r{2.28}.
Then $j$ as a left module is generated by matrix elements of
\be (\jb^{\ot n}-S_n)(\te p^{\hs\tu n}-r_n).\e{2.57a}\ee\ep

\bd In virtue of Theorem \re{t2.7}, \r{2.33} is an $\api$-bimodule 
and as a left module it is generated by matrix elements of 
$(\jb^{\ot 2}-R)\te p^{\hs\tu 2}-r$. Therefore
$j$ is the left module generated by 
\be (\jb^{\ot m}-R_k)\te p^{\hs\tu m}-r_{mk}, \e{2.57a'}\ee
$m=2,3,\k$, $k=1,\k,m-1$. We set $j_n$ as the left module generated by 
\r{2.57a'} for $m=2,\k,n$ (for $n=2$ it coincides with the old definition of 
$j_2$). Thus $\api j_n\subset j_n,\quad j_n\api\subset j_n,\quad
j_n\te p_i\subset j_{n+1},\quad \te p_i j_n\subset j_{n+1}$. Moreover,
\[ \te p^{\hs\tu n}\equiv R_k\te p^{\hs\tu n}+r_{nk}\quad (\mod.  j), \]
\[ \te p^{\hs\tu n}\equiv R_{i_1}(R_{i_2}\te p^{\hs\tu n}+r_{ni_2})+r_{ni_1}
\quad (mod.j),
\quad \mbox{ etc. }, \]
\[ \te p^{\hs\tu n}\equiv R_{\pi}\te p^{\hs\tu n}+r_{n\pi} \quad (mod.j),\quad
\te p^{\hs\tu n}\equiv S_n\te p^{\hs\tu n}+r_n \quad (mod.j). \]
For any minimal decomposition \r{2.56} we set $r_{n\pi}^{(i)}$ as the right 
hand side of \r{2.57}, where $i=(i_1,\k,i_s)$. We shall prove
\be r_{n\pi}^{(i)}\equiv r_{n\pi}^{(i')}\ (\mod. j_{n-1})\e{2.58}\ee
for any two such decompositions $i,i'$. But $i,i'$ can be obtained one from
another by a finite number of steps of the following 2 kinds:
\btr
\item[(i)] we replace $\k t_kt_l\k$ by $\k t_lt_k\k$ for $|k-l|>1$,
\item[(ii)] we replace $\k t_kt_{k+1}t_k\k$ by $\k t_{k+1}t_kt_{k+1}\k$.
\etr
Thus it suffices to check \r{2.58} for each of these two cases.

\noindent ad (i). We may assume $k<l-1$. One has
\[ (\jb^{\ot n}-R_k)r_{nl}=
(\jb^{\ot n}-R_k)p^{\tu(l-1)}\tp r\tp p^{\tu(n-l-1)}\equiv \]
\[ p^{\tu(k-1)}\tp r\tp p^{\tu(l-k-2)}\tp r\tp p^{\tu(n-l-1)}\equiv
(\jb^{\ot n}-R_l)r_{nk}\quad (\mod. j_{n-1}). \]
Thus $r_{nk}+R_kr_{nl}\equiv r_{nl}+R_lr_{nk}\quad
 (\mod. j_{n-1})$ and \r{2.58}
follows. 

\noindent ad (ii). In virtue of \r{2.41}
\[ r_{nk}+R_kr_{n,k+1}+R_kR_{k+1}r_{nk}\equiv \]
\[ r_{n,k+1}+R_{k+1}r_{nk}+R_{k+1}R_kr_{n,k+1}\quad (\mod. j_{n-1}) \]
and \r{2.58} follows also in this case. 

Thus in formula \r{2.57} we can use any minimal decomposition \r{2.56} for
computations modulo $j_{n-1}$. We shall prove
\be R_kr_n\equiv r_n-r_{nk}\quad (\mod. j_{n-1}),\quad n=2,3,\k.\e{2.59}\ee
Let $\pi\in\Pi_n$ be such that $\pi^{-1}(k)<\pi^{-1}(k+1)$ and \r{2.56} be a
minimal decomposition. Then $\pi'=t_k\pi$ satisfies 
${\pi\pri}^{-1}(k)>{\pi\pri}^{-1}(k+1)$ and has minimal decomposition 
$\pi'=t_kt_{i_1}\cdot\k\cdot t_{i_s}$. In such a way we get all $\pi'$ such
that ${\pi\pri}^{-1}(k)>{\pi\pri}^{-1}(k+1)$, each one exactly once. Due
to 
\r{2.57} and \r{2.58},
\be r_{n\pi'}\equiv R_kr_{n\pi}+r_{nk}\quad (\mod. j_{n-1}),\e{2.59'}\ee
$R_{\pi'}=R_kR_{\pi}$. Multiplying both sides by $R_k-\jb^{\ot n}$ and using 
\be R_kr_{nk}=-r_{nk}\e{2.60}\ee
(it follows from $Rr=-r$), we get 
\[(R_k-\jb^{\ot n})(r_{n\pi}+r_{n\pi'})\equiv-2r_{nk}\quad (\mod. j_{n-1}), \] 
$(R_k-\jb^{\ot n})(R_{\pi}+R_{\pi'})=0$. Thus 
\[ (R_k-\jb^{\ot n})r_n=\frac{1}{n!}\sum_{\pi} 
(R_k-\jb^{\ot n})(r_{n\pi}+r_{n\pi'})\equiv -r_{nk}\quad (mod. j_{n-1}) \] 
and \r{2.59} is proved. Moreover, 
\be (R_k-\jb^{\ot n})S_n=\frac{1}{n!}\sum_{\pi} (R_k-\jb^{\ot n})
(R_{\pi}+R_{\pi'})=0. \e{2.60a}\ee
Thus $S_n^2=\frac{1}{n!}\sum_{\pi}R_{\pi}S_n=\frac{1}{n!}\sum_{\pi}S_n=S_n$.
Using \r{2.59}, \r{2.59'} and mathematical induction w.r.t. the number of 
transpositions in a
minimal decomposition of $\pi$, one gets 
$R_{\pi}r_n\equiv r_n-r_{n\pi}\ (mod.j_{n-1})$.
Therefore 
\be S_nr_n\equiv r_n-\frac{1}{n!}\sum_{\pi}r_{n\pi}=r_n-r_n=0 
\quad(\mod. j_{n-1}).
\e{2.61} \ee
Using  
$S_n\te p^{\hs\tu n}\equiv \te p^{\hs\tu n}-r_n\ (\mod. j)$ and
\r{2.61}, 
$(\jb^{\ot n}-S_n)\te p^{\hs\tu n}\equiv r_n\equiv(\jb^{\ot n}-S_n)r_n
(\mod. j)$. Thus the elements \r{2.57a} belong to $j$. 

Let $\te j$ be the left module generated by \r{2.57a}. Then $\te j\subset j$.
We shall prove by mathematical induction that $j_n\subset\te j$. It is true 
for $n=2$ since
\be (\jb^{\ot 2}-R)\te p^{\hs\tu 2}-r=2(\jb^{\ot 2}-S_2)(\te p^{\hs\tu 2}-r_2)
\e{2.61'}\ee
(we use $S_2=\frac12+\frac12 R$, $r_2=\frac12 r$, $S_2r_2=0$).
If it is true for $n-1$ then using \r{2.59} and \r{2.60a}, we get
\[ (\jb^{\ot n}-R_k)\te p^{\hs\tu n}-r_{nk}\equiv 
(\jb^{\ot n}-R_k)(\te p^{\hs\tu n}-r_n)= \]
\[ (R_k-\jb^{\ot n})(S_n-\jb^{\ot n})(\te p^{\hs\tu n}-r_n)
\equiv 0 \quad (mod. \te j) \]
and $j_n\subset \te j$. Therefore $j\subset\te j$,  $j=\te j$.\ed

We set $S_0=\id_{\cb}$. 
Let $\al'=\{{\al\pri}_{in}:\ i=1,\k,\dim S_n\}$ be a basis of $\im\ S_n$,
$\beta'=\{{\beta\pri}_{jn}:\ j=1,\k,\dim(\jb^{\ot n}-S_n)\}$ be a basis of
$\im(\jb^{\ot n}-S_n)$. Then $\al'\sqcup\beta'$ is a basis of
$(\cb^{|\spi|})^{\ot n}$. We denote by $\al\sqcup\beta$ the dual basis. In
particular, 
\be \al^{in}(\jb^{\ot n}-S_n)=0. \e{2.61a} \ee
We set $\bpi'=\tb/j$ and ${p\pri}_k=\lam(\te p_k)$ where $\lam:\tb\lra\bpi'$ is
the canonical mapping.

\bc\e{c2.9} Let $K_N=\{\beta^{in}(\te p^{\hs\tu n}-r_n):\ i=1,\k,
\dim(\jb^{\ot n}-S_n),\ n=2,3,\k,N\}$, 
${L\pri}_N=\{\al^{in}({p\pri}^{\tu n}):\ i=1,\k,\dim\ S_n,\ n=0,1,2,\k,N\}$.
Then $K_{\infty}$ is a basis of $j$, ${L\pri}_{\infty}$ is a basis of $\bpi'$,
$K_N$ is a basis of $j^N=j\cap\tb^N$, ${L\pri}_N$ is a basis of 
\[ {\bpi\pri}^N=\api\cdot\spa\{{p\pri}_{i_1}\cdot\k\cdot{p\pri}_{i_n}:\ %
i_1,\k,i_n=1,\k,|\spi|,\ n=0,1,\k,N\} \] 
(we treat $j,j^N,{\bpi\pri},{\bpi\pri}^N$ as the left modules).\ec

\bd 
\be \jb^{\ot n}-S_n=\sum_i {\beta\pri}_{in}\beta^{in},\qquad 
\beta^{jn}(\jb^{\ot n}-S_n)=\beta^{jn},\e{2.61a'}\ee
 hence $j$ is the left module
generated by $K_{\infty}$. On the other hand, a finite combination 
$\sum a_{in}\beta^{in}(\te p^{\hs\tu n}-r_n)$, $a_{in}\in\api$, belongs to
$\tb^N$ iff $a_{in}=0$ for $n>N$ (Lemma \re{l2.1} and linear independence of
$\beta^{in}$ for given $n$). Therefore $j^N$ is generated by $K_N$ and 
(taking
$N=0$) elements of $K_{\infty}$ are linearly independent over $\api$. Hence,
$K_{\infty}$ is a basis of $j$, $K_N$ is a basis of $j^N$. Using 
$\jb^{\ot n}=\sum {\al\pri}_{in}\al^{in}+\sum {\beta\pri}_{in}\beta^{in}$,
$K_N\sqcup\{\al^{in}(\te p)^{\tu n}:\ i=1,2,\k,\dim\ S_n,\ n=0,1,2,\k,N\}$ is a
basis of $\tb^N$, $N\in\nb\cup\{\infty\}$ ($\tb^{\infty}=\tb$). Thus
${\bpi\pri}^N=\tb^N/j=\tb^N/j^N$ has a basis ${L\pri}_N$, 
$N\in\nb\cup\{\infty\}$ (${\bpi\pri}^{\infty}=\bpi\pri$).\ed

\bc\e{c2.9a}The left module $j^N=j\cap\tb^N$ is generated by \r{2.57a} for
$n=2,3,\k,N$. In particular, $j\cap\tb^1=\{0\}$, $j\cap\tb^2=j_2$.\ec

\bd It follows from Corollary \re{c2.9}, \r{2.61a'} and \r{2.61'}.\ed

\bp\e{p2.10} With the assumptions of Theorem \re{t2.8},
if $J$ satisfies the condition i) of Theorem \re{t2.8} and $J\cap\tb^2=j_2$
then 
$J=j$.\ep

\bd Clearly $j\subset J$. 
Let $J'=J/j\subset\bpi'$ and $N$ be the minimal number such that
${J\pri}^N=J'\cap {\bpi\pri}^N\neq\{0\}$. Therefore $N\geq3$, $\de
{J\pri}^N\subset {J\pri}^N\ot\api+\api\ot{J\pri}^N$. Let 
$0\neq x\in{J\pri}^N$. Then 
\[ x=\sum_{\ba{c} i=1,\k,\dim\ S_n\\n\leq N\ea} 
a_{in}\al^{in}{p\pri}^{\tu n}. \]
We set ${\bpi\pri}_n=\api\cdot\spa\{\al^{in}{p\pri}^{\tu n}:
\ i=1,2,\k,\dim\ S_n\}$, $\bpi\pri=\oplus_{n=o}^{\infty} {\bpi\pri}_n$.
The component of $\de x$ belonging to ${\bpi\pri}_{N-1}\ot{\bpi\pri}_1$
equals $0$. Thus 
\[ 0\equiv\sum_{i=1,\k,\dim S_N} \de(a_{iN})\al^{iN}_{j_1\k j_N}\times \]
\[ \sum_{k=1}^N {p\pri}_{j_1}\cdot\k\cdot{p\pri}_{j_{k-1}}
\la_{j_km}{p\pri}_{j_{k+1}}\cdot\k\cdot{p\pri}_{j_N}\ot{p\pri}_m
(\mod.\ {\bpi\pri}^{N-2}\ot{\bpi\pri}_1). \] 
In short,
\[ 0\equiv\sum_i\de(a_{iN})\sum_{k=1}^N\al^{iN}
\left[({p\pri}^{\tu(k-1)}\tp\la\tp{p\pri}^{\tu(N-k)})\ti p\pri\right].
\]
Since \r{2.4} holds in $\tb$, $p'\tp\la\equiv R(\la\tp p')\ (\mod.\api)$.
Using \r{2.61a}, $\al^{iN}=\al^{iN}S_N=\al^{iN}S_NR_k=\al^{iN}R_k$ and all
components in the second sum are equal modulo 
${\bpi\pri}^{N-2}\ot{\bpi\pri}_1$. We get
\[ 0\equiv\sum_i\de(a_{iN})\al^{iN}[(p'\tp\k\tp p'\tp\la)\ot p'], \]
\[ 0\equiv\sum_i a_{iN}^{(1)}\al^{iN}_{j_1\k j_N}
{p\pri}_{j_1}\k {p\pri}_{j_{N-1}}\la_{j_Nm}\ot a_{iN}^{(2)}\quad
(\mod.\ {\bpi\pri}^{N-2}\ot\api). \]
Acting by $\id\ot\eps$ and multiplying by $\la^{-1}_{mk}{p\pri}_k$, one has
\[ 0\equiv\sum_i a_{iN}\al^{iN}{p\pri}^{\tu N}\quad 
(\mod.\ {\bpi\pri}^{N-1}). \] 
Using Corollary \re{c2.9}, one obtains $a_{iN}=0$, $i=1,\k,\dim\ S_N$, 
$x\in J\pri\cap{\bpi\pri}^{N-1}=\{0\}$, contradiction. 
We get $J'=\{0\}$, $J=j$.\ed

\noindent {\em End of proof of Theorem \re{t2.8} }. We use the above facts.
\hfill$\Box$.\\

\bc\e{c2.11} Let $L_N=
\{\al^{in}p^{\tu n}:\ i=1,2,\k,\dim\ S_n,\ n=0,1,2,\k,N\}$.
Then $L_{\infty}$ is a basis of $\bpi$, $L_N$ is a basis of 
$\bpi^N=\api\cdot\spa\{p_{i_1}\k p_{i_n}:\ i_1,\k,i_n\in\spi,\ n=0,1,\k,N\}$ 
(we treat $\bpi$, $\bpi^N$ as left modules). In particular,
\be \dim_{\api} \bpi^N=\sum_{n=0}^N \dim\ S_n. \e{2.62}\ee\ec

\br\e{r2.99} Assume that the condition c. doesn't hold. Then we introduce
$\eta_i\in\api'$ as before, so \r{2.13} holds. But on the right hand side of
\r{2.15} (for $a=v_{AC}$, $b=w_{BD}$, $v,w\in\Irr\ H$)
we must add $L^{vw}_{mAB,CD}$ where 
$L^{vw}\in\mor(v\tp w,\la\tp v\tp w)$
and we don't have \r{2.16}, \r{2.31a}. Nevertheless, \r{2.6} is valid.
Therefore we get admissible $\eta_i$ in the following way. Let matrix 
elements
of nontrivial irreducible
representations $\{w^m\}_{m\in M}$ generate $\api$
as algebra with $I$. We put $\eta_i(I)=0$, assume some values of
$\eta_i(w^m_{AB})$ and using \r{2.13} compute $\phi_i(w^m_{AB})$. Then we 
have
the condition that $\rho$ of \r{2.6} preserves all the relations among
$w^m_{AB}$. We choose $\eta_i(\la_{js})$ so that \r{2.22} 
is
satisfied ($\eta_i$ are in general not determined uniquely by $\phi_i$). We 
also have an
additional condition that \r{2.31a'} holds 
for $b$ being matrix elements of $\{w^m\}_{m\in M}$
(due to Remark \re{r2.220} it implies \r{2.31a'} for all $b\in\api$ and 
using it we get $A_{\al r}=0$
and bimodule condition for $j_2$ in Theorem \re{t2.7}). 
Proposition \re{p2.14} holds provided $Mor(1,\la)=\{0\}$,
$Mor(\la,\la\tp\la)=\{0\}$ (otherwise we replace
$Mor(\ppi\tp\ppi,\ppi\tp\ppi)$ by $V_0$ in both places where $V_0$ is some
linear subspace of $Mor(\ppi\tp\ppi,\ppi\tp\ppi)$).
Moreover, in Proposition \re{p2.14}.2.c there is 
$B[Z\ot\jb+(R\ot\jb)(\jb\ot Z)]m$ on the right hand side of \r{2.81}
and also one more condition \r{2.82}. We don't get Proposition \re{p3.5}.2,
Proposition \re{p3.6} and Corollary \re{c3.7}. In Proposition \re{p3.5}.3,
Proposition \re{p3a.3}, Theorem \re{t3a.5} and Proposition 
\re{p3a.6} we assume
that $\la$ is a nontrivial representation. 
With such 
corrections,
the results of Sections 2--5 are still valid.\er

\bc\e{c2.13} 
\btr
\item[a)] $\bpi$ is the universal 
unital algebra generated by $\api$ and $p_i$ 
($i\in\spi$)
with the relations $I_{\bpi}=I_{\api}$,
\be p_sa=(a*f_{st})p_t+a*\eta_s-\la_{st}(\eta_t*a),\qquad a\in\api,\e{2.63}\ee
\be 
(R-\jb^{\ot 2})_{kl,ij}
(p_ip_j-\eta_i(\la_{js})p_s+T_{ij}-\la_{im}\la_{jn}T_{mn})=0.
\e{2.64}\ee 
\item[b)] $\bpi$ is the 
universal unital algebra generated by $\api$ and $p_1,\k,p_s$
with the relations $I_{\bpi}=I_{\api}$, 
\be (\ppi\tp w)N_w=N_w(w\tp\ppi),\quad w\in \Rep\ H,\e{2.65}\ee
\be R_P(\ppi\tp\ppi)=(\ppi\tp\ppi)R_P, \e{2.66}\ee
where $\ppi$ is given by the condition 3., 
\be R_P=\left(\ba{cccc} R & Z & -R\cdot Z & (R-\jb^{\ot 2})T \\
                        0 & 0 &      \jb  &     0      \\
			0 &\jb&      0    &     0      \\
			0 & 0 &      0    &     1      \ea\right),
N_w=\left(\ba{cc} G_w & H_w \\ 0 & \jb_w \ea\right), 
\e{1.18mm}\ee
$(G_w)_{iC,Dj}=f_{ij}(w_{CD})$, $(H_w)_{iC,D}=\eta_i(w_{CD})$, $R=G_{\la}$,
$Z=H_{\la}$.\etr\ec

\br\e{r2.uff} For $\eta=0$, $T=0$ 
(that choice always satisfies the conditions \r{2.28}, \r{2.31b}, \r{2.32})
we get $p_s\la_{ij}=R_{si,mt}\la_{mj}p_t$,
$R_{kl,ij}p_ip_j=p_kp_l$ (cf \cite{S}, \cite{M}).\er

\br\e{r2.uf} In \r{2.63} it suffices to take $a$ being
generators of $\api$
(as algebra with unity), 
in \r{2.65} it suffices to take $\{w^m\}_{m\in M}\subset \Rep\ H$
such that matrix elements of $w^m$ generate $\api$.\er

\br\e{r2.103} Replacing $T$ by 
$T'=\frac12(\jb^{\ot 2}-R)T$, we don't change \r{2.64}.
One has $RT'=-T'$. So in the following we can (and will) assume 
\be RT=-T. \e{2.67}\ee\er

\bd a) follows from Theorem \re{t2.8}. Due to \r{2.13} and \r{2.5} it 
suffices
to take $a$ as generators.

ad b) Let $a=w_{mn}$, $m,n=1,\k,\dim w$. Then \r{2.3} implies 
\[ (\la\tp w)G_w=G_w(w\tp\la), \]
\r{2.63} is equivalent to 
\be p\tp w=G_w(w\tp p)+H_ww-(\la\tp w)H_w. \e{2.69a}\ee
We can rewrite these two equations as \r{2.65}. One can replace \r{2.64} by
\be (R-\jb^{\ot 2})(p\tp p-Z\cdot p+T-(\la\tp\la)T)=0.\e{2.68'}\ee
Using \r{2.69a} for $w=\la$, this is equivalent to \r{2.66}.\ed

\bp\e{p2.14H} $\bpi$ is a Hopf algebra (with invertible coinverse).\ep

\bd Let $w,w'\in \Irr\ H$. Then 
\[ w\tp w'\simeq\op_{w''\in \Irr H}\ c^{w''}_{ww'}w'' \]
for some $c^{w''}_{ww'}\in\nb$.
Thus there exist linearly independent $S^{\al}_{ww'w''}\in\mor(w'',w\tp w')$,
$\al=1,\k,c^{w''}_{ww'}$. Then $\api$ is the algebra generated by 
matrix elements of unequivalent irreducible representations of $H$ satisfying
\be (w\tp w')S^{\al}_{ww'w''}=S^{\al}_{ww'w''}w'',\quad w,w',w''\in\Irr\ H,
\quad \al=1,\k,c^{w''}_{ww'}.\e{2.68''}\ee
We conclude that $\bpi$ is the universal algebra generated by (matrix
elements of) a set
of  representations of $G$ ($\ppi$ and $w\in \Irr\ H$) 
satisfying \r{2.65}, \r{2.66}, \r{2.68''},
\[ \ppi=I\tp\ppi, \quad \ppi i=i\la \mbox{ and } s\ppi=Is \]
where $i:\cb^{|\spi|}\lra\cb^{|\spi|}\op\cb,
\quad s:\cb^{|\spi|}\op\cb\lra\cb$ are the canonical mappings, 
$I$ is the trivial representation of $H$.
Thus the relations are given by
morphisms. Moreover, these representations are invertible:
\[ \ppi^{-1}=\left(\ba{cc} \la^{-1} & -\la^{-1}p \\ 0 & I \ea\right), \]
$w^{-1}=S(w)$ for $w\in \Irr\ H$. Using the arguments of \cite{P} 
or \cite{WZ}, we get that $\bpi$ 
has a coinverse $S$. Similarly, ${\cal P}^T$ and $w^T$,
$w\in\mbox{ Irr }H$, are invertible representations of
$G^{\mbox{opp}}$, where $\mbox{ Poly }(G^{\mbox{opp}})=
({\cal B},\tau\Delta)$  (coinverse of ${\cal A}$ is
invertible). Hence $({\cal B},\tau\Delta)$ has a coinverse $S'$,
by the general theory $S'=S^{-1}$.\ed

 Let
\be \tau^{kl}=(R-\jb^{\ot 2})_{kl,ij}\tau_{ij}, \e{2.68}\ee
where $\tau_{ij}$ are defined in Theorem \re{t2.7}.

\bp\e{p2.14'}

\mbox{}
\btr
\item[1)] \r{2.28} is equivalent to
\be (\la\tp\la)_{kl,ij}(\tau^{ij}*b)=b*\tau^{kl},\quad b\in S,\e{2.69}\ee
where $S$ is a set generating $\api$ as an algebra with unity
\item[2)] 
\[ \tau^{ij}(ab)=(f_{js}*f_{im})(a)\tau^{ms}(b)+\tau^{ij}(a)\eps(b)\quad
(a,b\in\api),\quad \tau^{ij}(I)=0. \]
\etr\ep

\bd 1) follows from Remark \re{r2.220}, \r{2.17g}, \r{2.34a} and \r{2.17a}.
2) follows from Proposition \re{p2.7'}, \r{2.29}, \r{2.34a} and
\[ R_{kl,ij}(f_{js}*f_{im})=(f_{lj}*f_{ki})R_{ij,ms} \]
( we get it acting $f_{ij}$ on \r{2.3}).\ed

\bp\e{p2.14} Let $R\not\in\cb\jb^{\ot 2}$. One has
\btr
\item[1)] $\mor(\ppi\tp\ppi,\ppi\tp\ppi)=\cb\cdot\id\op\cb R_P\op
\{m_P:(\la\tp\la)m=m\}$ where $R_P$ is given by \r{1.18mm} and
\be m_P=\left(\ba{cccc}     0 & 0 & 0 & m\\
                            0 & 0 & 0 & 0\\
	                    0 & 0 & 0 & 0\\
			    0 & 0 & 0 & 0\ea\right).\e{1.18m}\ee

\item[2)] 
$W\in\mor(\ppi\tp\ppi,\ppi\tp\ppi)$ satisfies 
			
\be (W\ot\jb)(\jb\ot W)(W\ot\jb)=(\jb\ot W)(W\ot\jb)(\jb\ot W)\e{1.18a}\ee
if and only if 
\itemite a) $W=x\cdot\id$ ($x\in\cb\setminus\{0\}$) or
\itemite b) $W\in\{m_P:\ (\la\tp\la)m=m\}$ or
\itemite c) $W=y\cdot(R_P+m_P)$ for $y\in\cb\setminus\{0\}$ and $m$ such that
 $(\la\tp\la)m=m$
provided that
\be [(R-\jb^{\ot 2})\ot\jb]F=0,\e{2.80}\ee
\be A_3(Z\ot\jb-\jb\ot Z)T=0.\e{2.81}\ee

Those $W$ are invertible if and only if we have the case a) or c).\etr\ep

\br Examples of R-matrices for inhomogeneous quantum groups were given e.g. in
\cite{Cel}, \cite{S}, \cite{ChD}, \cite{L}.\er

\bd ad 1) One has
\[ \ppi\tp\ppi=\left(\ba{cccc} \la\tp\la & \la\tp p & p\tp\la & p\tp p \\
                                   0     &    \la   &    0    &   p    \\
				   0     &     0    &   \la   &   p    \\
				   0     &     0    &    0    &   I
\ea\right). \]
We assume
\[ W=\left(\ba{cccc} A&B&E&F\\C&D&G&H\\J&K&N&P\\L&M&Q&U \ea\right)
\in\mor(\ppi\tp\ppi,\ppi\tp\ppi). \]
It gives a set of linear relations on matrices $A,B,\k,U$. Using \r{2.68'} and
\be p\tp\la=R(\la\tp p)+Z\la-(\la\tp\la)Z \e{2.69'}\ee
(it is \r{2.69a} for $w=\la$) one can solve them and get
$A=b+aR$, $B=aZ$, $E=-aRZ$, $F=a(R-\jb^{\ot 2})T+m$,
$D=N=b\jb$, $G=K=a\jb$, $U=a+b$, $C=H=J=P=L=M=Q=0$, where $a,b,k\in\cb$, 
$(\la\tp\la)m=m$. It means $W=b\cdot\id+a\cdot R_P+m_P$ and 1) follows.

ad 2) We set $l=Z\ot\jb+(R\ot\jb)(\jb\ot Z)$.
Using \r{2.15}, we get $l_{ijk,rs}=\eta_i((\la\tp\la)_{jk,rs})$ and 
(see \r{2.17a})
\be (\jb\ot R)l=lR. \e{2.82}\ee
Moreover, $(\la\tp\la)m=m$ gives 
\be lm=0 \e{2.83}\ee
and (acting by $f_{ij}$)
\be (R\ot\jb)(\jb\ot R)(m\ot\jb)=\jb\ot m,\e{2.84}\ee
hence (using $R^2=\jb^{\ot 2}$)
\be (\jb\ot R)(R\ot\jb)(\jb\ot m)=m\ot\jb.\e{2.85}\ee

We shall check for which $m$ and $x,y\in\cb$, 
$W=x\cdot\id+y\cdot R_P+m_P$
satisfies \r{1.18a}. Since $\ppi$ acts in $\cb^{|\spi|}\op\cb$, $W$ acts on 
\[ \cb^{|\spi|}\ot\cb^{|\spi|}\op\cb^{|\spi|}\ot\cb\op\cb\ot\cb^{|\spi|}\op
\cb\ot\cb. \]
Denoting the standard basis elements in 
$\cb^{|\spi|}\op\cb$ by $e_i$ ($i\in\spi$)
and $f$, one gets 
\be\left.\ba{rcl}
R_P(e_i\ot e_j) & = & R_{kl,ij}e_k\ot e_l,\\\\
R_P(e_i\ot f)   & = & Z_{kl,i}e_k\ot e_l+f\ot e_i,\\\\
R_P(f\ot e_i)   & = & -(RZ)_{kl,i}e_k\ot e_l+e_i\ot f,\\\\
R_P(f\ot f)     & = & ((R-\jb^{\ot 2})T)_{ij}e_i\ot e_j+f\ot f,\\\\
m_P(e_i\ot e_j) & = & 0,\\\\
m_P(e_i\ot f)   & = & 0,\\\\
m_P(f\ot e_i)   & = & 0,\\\\
m_P(f\ot f)     & = & m_{ij}e_i\ot e_j.\ea\right\}\e{2.70}\ee

Let us restrict ourselves to
$\cb^{|\spi|}\ot\cb^{|\spi|}\ot\cb^{|\spi|}$. Then \r{1.18a} gives an 
analogous 
formula for $x\cdot\jb^{\ot 2}+y\cdot R$. Using \r{2.40} and $R^2=\jb^{\ot 2}$,
it
means $x^2y(R\ot\jb-\jb\ot R)=0$, $x=0$ or $y=0$ ($R\ot\jb=\jb\ot R$ would
mean $V\ot\cb^4=\cb^4\ot V$ where $V=\ker(R+\jb^{\ot 2})$, 
$R\in\cb\jb^{\ot 2}$,
contradiction). 

Setting $x\neq 0$, $y=0$ and applying both sides of \r{1.18a} to 
$f\ot f\ot e_k$, one
obtains $m_{ij}(e_i\ot e_j\ot e_k)=0$, $m=0$. Clearly
$x\neq 0$, $y=0$, $m=0$ gives a
solution of \r{1.18a}. The same holds for $x=y=0$ (both sides of \r{1.18a}
equal $0$). It remains to consider $W=y(R_P+m_P)$ for $y\neq0$. In order
to check \r{1.18a} we may assume $y=1$.  
Using \r{2.70}, we
find that \r{1.18a} on $e_i\ot e_j\ot e_k$ follows from \r{2.40}, on
$e_i\ot e_j\ot f$, $e_i\ot f\ot e_j$, $f\ot e_i\ot e_j$ is equivalent to
\r{2.82}, on $e_i\ot f\ot f$, $f\ot e_i\ot f$, $f\ot f\ot e_i$ is equivalent
to \r{2.80} (we use \r{2.55a}, \r{2.82} and \r{2.85}),
on $f\ot
f\ot f$ is equivalent to $Bls=0$ where 
$B$ is given by \r{2.43}, $s=s_0+m$,
$s_0=(R-\jb^{\ot 2})T=-2T$ 
(see \r{2.67}).
Using \r{2.83} and $[\jb\ot(\jb^{\ot 2}+R)]ls_0=0$
(which follows from \r{2.82}), we get
\[ Bls=Bls_0=\frac12A_3ls_0=-A_3(Z\ot\jb-\jb\ot Z)T \]
and $Bls=0$ is equivalent to \r{2.81}. 

Invertibility condition is obvious (in the case c) we use the existence
of $R_P^{-1}=R_P$).\ed

\br One can also consider the case when $(R+\jb^{\ot 2})(R-Q\jb^{\ot 2})=0$
where $Q\neq0,\pm1$ is not a root of unity. Then $(\rho+\id)(\rho-Q\id)=0$
and we should replace everywhere $R-\jb^{\ot 2}$ by $R-Q\jb^{\ot 2}$, 
$R_{kl,ij}p_ip_j=p_kp_l$ by $R_{kl,ij}p_ip_j=Qp_kp_l$,
\[ A_3=Q^3\jb\ot\jb\ot\jb-Q^2R\ot\jb-Q^2\jb\ot R+Q(R\ot\jb)(\jb\ot R)+ \]
\[ Q(\jb\ot R)(R\ot\jb)-(R\ot\jb)(\jb\ot R)(R\ot\jb), \]
\[ A=(R\ot\jb)(\jb\ot R)-Q(\jb\ot R)+Q^2\jb\ot\jb\ot\jb,\]
\[ B=(\jb\ot R)(R\ot\jb)-QR\ot\jb+Q^2\jb\ot\jb\ot\jb,\]
\[  R_P=\left(\ba{cccc} R & Z & (Q-1-R)\cdot Z & (R-Q\jb^{\ot 2})T \\
                        0 & 0 &      Q\jb  &     0      \\
			0 &\jb&      (Q-1)\jb    &     0      \\
			0 & 0 &      0    &     Q      \ea\right),\]
$n!$ is replaced by 
\[ (n)_Q!=\sum_{\pi\in\Pi_n} Q^{s(\pi)} = (1)_Q(2)_Q\cdot\ldots\cdot(n)_Q \]
where $s(\pi)$ is the number of transpositions in the minimal decomposition
of $\pi$ and $(k)_Q=1+Q+\ldots+Q^{k-1}$ (what concerns $S_n$ and $A_3$
see \cite{J}),
\[ r_{n\pi}=Q^{s-1}r_{ni_1}+Q^{s-2}R_{i_1}r_{ni_2}+\ldots+
R_{i_1}\cdot\ldots\cdot R_{i_{s-1}}r_{ni_s} \]
where $s=s(\pi)$, in Remark \re{r2.103} $T'=\frac1{1+Q}(Q\jb^{\ot 2}-R)T$.
In Proposition \re{p2.14}.2.c one has $W=y(R_P+m_P)^{\pm1}$,
$(\jb\ot R)(R\ot\jb)(\jb\ot m)-m\ot\jb$ on the right hand side of \r{2.80}
 and additional 
condition $Rm=-m$. In \r{3.4} one has $\frac1{(1+Q)c^2}(Q\jb^{\ot 2}-\hat R)$
instead of $\frac1{2c^2}(\jb^{\ot 2}-\hat R)$. In \r{3.8'} one obtains 
$R^{-1}_{ab,jl}$ on the left hand side. In Proposition \re{p3.5}.2,
Proposition \re{p3a.3}, Theorem \re{t3a.5} and Proposition \re{p3a.6}
we assume $|Q|=1$ (otherwise existence of the considered $*$-structure
in $\bpi$ would imply $R=-\jb^{\ot 2}$, $\bpi=\api\cdot\spa\{I,p_i\}$).
With these corrections, all the results (in particular Remark \re{r2.99})
remain true but we do not get Proposition \re{p3.6}, Corollary \re{c3.7}
and there are small modifications in the proofs.\er

\sectio{Isomorphisms and $*$ structure}

In this Section we consider isomorphisms among inhomogeneous quantum groups 
as
well as $*$-structures on them.
 Throughout the Section we assume that $\po(H)=(\api,\de)$ is a
Hopf algebra
satisfying the conditions a.--c. and $\po(G)=(\bpi,\de)$ is the corresponding
Hopf algebra as in Theorems \re{t2.5} and \re{t2.8}. Then $G$ is called an
inhomogeneous quantum group. 

\bp\e{p3.1} Let $w\in \Irr\ G$. Then $w\in \Irr\ H$. \ep

\bd Let $W=\spa\{w_{mn}:\ m,n=1,\k,\dim w\}$ and $s$ be the 
smallest natural number such
that $W\subset\bpi^s$ ($\bpi=\cup_s\bpi^s$). Assume that $s>0$. Then 
$\de\bpi^s\subset\bpi^{s-1}\ot\bpi^s+\bpi^s\ot\bpi^{s-1}$. There exists
$\phi\in(\bpi^s)'$ such that $\phi_{|_{\bpi^{s-1}}}=0$ and
$\phi_{|_{W}}\neq0$. Therefore $(\id\ot\phi)\de W\subset\bpi^{s-1}$, 
$w\phi(w)\in M_{\dim\ w}(\bpi^{s-1})$. Moreover, one can choose
$x\in\cb^{\dim\ w}$ such that $\phi(w)x\neq0$. We take $\phi(w)x$ as 
the first
vector of a basis in the carrier vector space of $w$. Thus
$w_{k1}\in\bpi^{s-1}$, $k=1,\k,\dim\ w$, 
$w_{kl}\ot w_{l1}=\de w_{k1}\in\bpi^{s-1}\ot\bpi^{s-1}$. Using linear
independence of $w_{l1}$ ($w\in \Irr\ G$), we get $w_{kl}\in\bpi^{s-1}$,
$W\subset\bpi^{s-1}$, contradiction. Thus $s=0$, $W\subset\bpi^0=\api$.\ed 

\bc\e{c3.2}
\[ \api=\spa\{w_{kl}:\ k,l=1,\k,\dim\ w,\ w\in \Irr\ G\}. \]
Thus $\bpi$ determines $\api$ uniquely.\ec

\bp\e{p3.3} Let $x\in\bpi$, $\de x\in\api\ot\bpi+\bpi\ot\api$. Then
$x\in\bpi^1$.\ep 

\bd Let $N$ be the minimal number such that $x\in\bpi^N$. Assume $N\geq 2$.
Then 
\[ x=\sum_{\ba{c} i=1,\k,\dim\ S_n\\n\leq N\ea} a_{in}\al^{in}p^{\tu n}. \]
Using the same arguments as in the proof of Proposition \re{p2.10}, one gets
$a_{iN}=0$, $i=1,\k,\dim\ S_N$, $x\in\bpi^{N-1}$, contradiction. Thus
$x\in\bpi^1$.\ed 

\bp\e{p3.4} One has
\mbox{}
\btr
\item[1)] Suppose that $\bpi,\api,\de,\la,p,f,\eta,T$ and 
$\hat\bpi,\hat\api,\hat\de,\hat\la,\hat p,\hat f,\hat\eta,\hat T$ describe 
two
inhomogeneous quantum groups $G,\hat G$ and 
$\phi:\bpi\lra\hat\bpi$ be an isomorphism of
bialgebras. Then $\phi(\api)=\hat\api$, $\phi(\bpi^1)=\hat\bpi^1$. We denote
$\phi_{\api}=\phi_{|_{\api}}:\api\lra\hat\api$. 
\item[2)] Moreover, let $\phi(\la)=M\hat\la M^{-1}$ for an invertible matrix
$M$. Then 
\be \phi(p)=M(c\hat p+h-\hat\la h), \e{3.1}\ee
for some $c\in\cb\setminus\{0\}$, $h_s\in\cb$ ($s\in\spi$) and we can choose 
\be \hat f=(M^{-1}fM)\circ\phi_{\api}^{-1},\e{3.2}\ee
\be \hat\eta=\frac1c M^{-1}(\eta+fMh-\eps Mh)\circ\phi_{\api}^{-1},\e{3.3}\ee
\be \hat T=\frac1{c^2}(M^{-1}\ot M^{-1})T
+\frac1{2c^2}(\jb^{\ot 2}-\hat R)[-(M^{-1}\ot M^{-1})ZMh+h\ot h]
\e{3.4}\ee
where $Z=\eta(\la)$, 
\be \hat R=(M^{-1}\ot M^{-1})R(M\ot M).\e{3.6}\ee
\item[3)] Let $\bpi,\api,\de,\la,p,f,\eta,T$ describe an inhomogeneous 
quantum
group, $\hat\api,\hat\de,\hat\la$ satisfy the conditions a.--c.,
$\phi_{\api}:\api\lra\hat\api$ be an isomorphism of bialgebras such that
$\phi_{\api}(\la)=M\hat\la M^{-1}$ for an invertible matrix $M$ and
$c\in\cb\setminus\{0\}$, $h_s\in\cb$ ($s\in\spi$). 
Then there exists an inhomogeneous quantum group
described by $\hat\bpi,\hat\api,\hat\de,\hat\la,\hat p,\hat f,\hat\eta,
\hat T$
and isomorphism of bialgebras $\phi:\bpi\lra\hat\bpi$ such that
$\phi_{\api}=\phi_{|_{\api}}$ and \r{3.1}--\r{3.4} hold.\etr\ep

\bd ad 1) According to Corollary \re{c3.2}, 
\[ \phi(\api)=\spa\{ \phi(w_{kl}):\ k,l=1,\k,\dim w,\ w\in \Irr\ G\}= \]
\[ \spa\{ v_{kl}:\ k,l=1,\k,\dim v,\ v\in \Irr\ \hat G\}=\hat\api. \]
Let $x\in\bpi^1$. Then $\de x\in\bpi\ot\api+\api\ot\bpi$, 
$\de\phi(x)=(\phi\ot\phi)\de x\in\hat\bpi\ot\hat\api+\hat\api\ot\hat\bpi$.
Using Proposition \re{p3.3}, we get $\phi(x)\in\hat\bpi^1$. Thus
$\phi(\bpi^1)\subset\hat\bpi^1$. Interchanging $\bpi$ with $\hat\bpi$, one gets
$\phi(\bpi^1)=\hat\bpi^1$. 

\noindent ad 2) $\phi(p)=k\cdot\hat p+l$ for some $k_{ij},l_i\in\hat\api$.
Therefore 
\[ (k\hat p+l)\ti I+M\hat\la M^{-1}\ti(k\hat p+l)=
(\phi\ot\phi)(p\ti I+\la\ti p)= \]
\[ (\phi\ot\phi)\de p=\hat\de\phi(p)=
\hat\de(k)(\hat p\ti I+\hat\la\ti\hat p)+\hat\de(l). \]
We get $\hat\de(k)=k\ti I$, $k_{ij}\in\cb$, $M\hat\la M^{-1}k=k\hat\la$, 
$l\ti I+M\hat\la M^{-1}\ti l=\hat\de(l)$. 
Thus $k=c\cdot M$ and (cf \r{2.21} and
later formulae) $l=\hat h-M\hat\la M^{-1}\hat h$ for some 
$c\in\cb\setminus\{0\}$, $\hat h_s\in\cb$
($s\in\spi$). Setting $h=M^{-1}\hat h$ one obtains \r{3.1}. 

Acting $\phi$ on the relation $pa=(a*f)p+a*\eta-\la(\eta*a)$, we get 
\be\left.\ba{l} M(c\hat p+h-\hat\la h)b= \\\\
(b*f\circ\phi_{\api}^{-1})M(c\hat p+h-\hat\la h)+\\\\
b*\eta\circ\phi_{\api}^{-1}-
M\hat\la M^{-1}(\eta\circ\phi_{\api}^{-1}*b),\ea\right\}\e{3.5}\ee
where $b=\phi(a)\in\hat\api$. But (acting $\phi$ on \r{2.3})
\[ (b*f\circ\phi_{\api}^{-1})M\hat\la M^{-1}=
M\hat\la M^{-1}(f\circ\phi_{\api}^{-1}*b). \]
Thus \r{3.5} is equivalent to
\[ \hat pb=\{b*[(M^{-1}fM)\circ\phi_{\api}^{-1}]\}\hat p+ \]
\[ \frac1c\{b*[M^{-1}(\eta+fMh-\eps Mh)\circ\phi_{\api}^{-1}]\}- \]
\[ \frac1c\hat\la\{[M^{-1}(\eta+fMh-\eps Mh)\circ\phi_{\api}^{-1}]*b\}. \]
It proves \r{3.2} and \r{3.3}. Applying \r{3.2} and \r{3.3} to $\hat\la$ one
obtains \r{3.6} and
\be \hat Z=\frac1c[(M^{-1}\ot M^{-1})ZM+\hat R(\jb\ot h)-h\ot\jb].\e{3.7}\ee
Acting $\phi$ on the relation \r{2.68'}
and using \r{3.6}, we get 
\be\left.\ba{c} (\hat R-\jb^{\ot 2})[\hat p\tp\hat p+\hat p\tp h'-
\hat p\tp\hat\la h'+ \\\\
 h'\tp\hat p-\hat\la h'\tp\hat p+\hat\la h'\tp\hat\la h'-\hat\la h'\tp h'- 
h'\tp\hat\la h'+h'\tp h'-\\\\
Z'\hat p-Z'h'+Z'\hat\la h'+T'-(\hat\la\tp\hat\la)T']=0,\ea\right\}\e{3.8}\ee
where $Z'=\frac1c(M^{-1}\ot M^{-1})ZM$, $T'=\frac1{c^2}(M^{-1}\ot M^{-1})T$, 
$h'=\frac{h}{c}$. But
\[ \hat p\tp\hat\la=
\hat R(\hat\la\tp\hat p)+\hat Z\hat\la-(\hat\la\tp\hat\la)\hat Z, \]
so \r{3.8} is equivalent to
\[ (\hat R-\jb^{\ot 2})[\hat p\tp\hat p-G\hat p+U]=0, \]
where
\[ G=-\jb\ot h'+\hat R(\hat\la h'\ot\jb)-h'\ot\jb+\hat\la h'\ot\jb+Z'= \]
\[ \hat Z+(\hat R+\jb^{\ot 2})(\hat\la h'\ot\jb-\jb\ot h'), \]
\[ U=-\hat Z\hat\la h'+(\hat\la\tp\hat\la)\hat Z h'+\hat\la h'\tp\hat\la h'-
\hat\la h'\tp h'-h'\tp\hat\la h' + \]
\[ h'\tp h'-Z'h'+Z'\hat\la h'+T'-(\hat\la\tp\hat\la)T'. \]
But
\[ Z'\hat\la h'-\hat Z \hat\la h'=h'\tp\hat\la h'-\hat R(\hat\la h'\tp h'), 
\]
\[ (\hat\la\tp\hat\la)\hat Z h'=(\hat\la\tp\hat\la)Z'h'+
\hat R(\hat\la\tp\hat\la)(h'\tp h')-(\hat\la\tp\hat\la)h'\tp h', \]
hence
\[ U=\te T-(\hat\la\tp\hat\la)\te T+
(\hat R+\jb^{\ot 2})(\hat\la h'\tp\hat\la h'-
\hat\la h'\tp h'), \]
where $\te T=T'-Z'h'+h'\tp h'$. Therefore \r{3.8} is equivalent to 
\[ (\hat R-\jb^{\ot 2})(\hat p\tp\hat p-\hat Z\hat p+\te T-
(\hat\la\tp\hat\la)\te T)=0.
\] 
Thus we can choose $\hat T$ as $\frac12(\jb^{\ot 2}-\hat R)\te T$ 
(cf Remark \re{r2.103}) and \r{3.4} follows.

ad 3) We define $\hat f,\hat\eta,\hat T$ by \r{3.2}--\r{3.4} and
$\hat\bpi$ as the 
universal algebra generated by $\hat\api$ and $\hat p_i$, $i\in\spi$,
satisfying $I_{\hat\bpi}=I_{\hat\api}$,
\[ \hat p b=(b*\hat f)\hat p+(b*\hat\eta)-\hat\la(\hat\eta*b),\quad 
b\in\hat\api, \]
\[ (\hat R-\jb^{\ot 2})[\hat p\tp\hat p-\hat Z\hat p+
\hat T-(\hat\la\tp\hat\la)\hat T]=0, \]
where $\hat f, \hat\eta, \hat T$ are given by \r{3.2}--\r{3.4},
$\hat R_{ij,kl}=\hat f_{il}(\hat\la_{jk})$, $\hat Z=\hat\eta(\hat\la)$.
The computations in 2) show that there exists 
a unital homomorphism of algebras
$\phi:\bpi\lra\hat\bpi$ such that $\phi_{|_{\api}}=\phi_{\api}$ and \r{3.1}
holds ($\phi$ transforms the relations in $\bpi$ into the relations in
$\hat\bpi$). The same computations show that there exists a unital 
homomorphism of algebras $\phi':\hat\bpi\lra\bpi$ such that 
${\phi\pri}_{|\ \hat\api}=\phi_{\api}^{-1}$ and
$\phi'[M(c\hat p+h-\hat\la h)]=p$, i.e.
$\phi'(\hat p)=\frac{1}{c}M^{-1}[p-Mh+\la Mh]$. 
Thus $\phi\phi'=\phi'\phi=\id$
and $\phi$ is an isomorphism. We set $\hat\de=(\phi\ot\phi)\de\phi^{-1}$.
Hence $(\hat\bpi,\hat\de)$ is a bialgebra with the proper bialgebra 
structure on $\hat\api$ and $\phi$ is an isomorphism of bialgebras. 
Computations in 2) show $\hat\de\hat p=\hat p\ti I+\hat\la\ti\hat p$ and
the properties of $(\bpi,\de)$ imply that $(\hat\bpi,\hat\de)$ corresponds to
an inhomogeneous quantum group described by
$\hat\bpi,\hat\api,\hat\de,\hat\la,\hat p,\hat f,\hat\eta,\hat T$.\ed

\bp\e{p3.5} Let $\bpi,\api,\de,\la,p,f,\eta,T$  correspond to an 
inhomogeneous
quantum group where $(\api,\de)$ is a Hopf 
${}^*$-algebra such that $\bar\la=\la$.
\btr
\item[1)] Let $(\bpi,\de)$ be a Hopf ${}^*$-algebra such that
$*_{|_{\api}}=*_{\api}$. Then there exist $m\neq0,n_s\in\cb$ ($s\in\spi$) such 
that
\be {p\pri}_i=mp_i+n_i-\la_{ij}n_j\e{3.4a}\ee
satisfy ${p\pri}_i^*={p\pri}_i$. In particular, there exist 
$\hat\bpi,\hat\api,\hat\de,\hat\la,\hat p,\hat f,\hat\eta,\hat T$
corresponding to an inhomogeneous quantum group and 
Hopf ${}^*$-algebra isomorphism
$\phi:\bpi\lra\hat\bpi$ such that $\api=\hat\api$, $\phi_{|_{\api}}=\id$,
$\phi({p\pri}_i)=\hat p_i$, $\hat p_i^*=\hat p_i$.
\item[2)] There exists Hopf ${}^*$-algebra structure in $\bpi$ such that
$*_{|_{\api}}=*_{\api}$ and $p_i^*=p_i$ ($i\in\spi$) iff 
\be f_{ij}(S(a^*))=\ov{f_{ij}(a)},\qquad a\in\api,\e{3.5'}\ee
\be \eta_i(S(a^*))=\ov{\eta_i(a)},\qquad a\in\api,\e{3.6'}\ee
\be \te T-T\in\mor(I,\la\tp\la),\e{3.6a}\ee
where $\te T_{ij}=\ov{T_{ji}}$, $i,j\in\spi$. Moreover, such $*$ is
unique. 
\item[3)] Proposition \re{p3.4} remains valid if we consider $(\api,\de)$,
$(\hat\api,\hat\de)$, $(\bpi,\de)$ and $(\hat\bpi,\hat\de)$ as 
Hopf ${}^*$-algebras,
$\phi,\phi_{\api}$ as Hopf ${}^*$-algebra isomorphisms, 
$p_i,\hat p_i$ as selfadjoints,
$M=\bar M$ and $c\in\rb\setminus\{0\}$, 
$h_i\in\rb$ ($i\in\spi$). Moreover, one has 
$\ov{\hat\la}=\hat\la$.\etr\ep

\br Statements 1) and 3) remain valid if we replace everywhere
Hopf ${}^*$-algebras by ${}^*$-bialgebras.\er

\br If $\la=\bar\la$, $\hat\la=\ov{\hat\la}$ and
$\phi_{\api}:\api\lra\hat\api$ is a ${}^*$-isomorphism such that
$\phi_{\api}(\la)=M\hat\la M^{-1}$ then we can assume $M=\bar M$ (Conjugating
one has $\phi_{\api}(\la)=\bar M\hat\la\bar M^{-1}$. Using the condition b.,
$\bar M=\al\cdot M$ for some $\al=e^{i\phi}$, $\phi\in\rb$. Replacing $M$ by
$M'=e^{i\phi/2}M$, one gets $\phi_{\api}(\la)=M'\hat\la{M\pri}^{-1}$ and
$M'=\ov{M'}$).\er 

\bd ad 1) Acting by $*$ on \r{2.0}, we get 
\[ \de p_i^*=\la_{ij}\ot p_j^*+p_i^*\ot I\in\api\ot\bpi+\bpi\ot\api. \]
Using Proposition \re{p3.3}, $p_i^*=k_{ij}p_j+l_i$ for some
$k_{ij},l_i\in\hat\api$. Therefore
\[ (kp+l)\ti I+\la\ti(kp+l)=\de(k)(p\ti I+\la\ti p)+\de(l). \]
We get (cf the proof of Proposition \re{p3.4}) $k=cI$, $l=g-\la g$ for some
$c,g_s\in\cb$ ($s\in\spi$). Thus $p_j^*=dp_j+g_j-\la_{jk}g_k$. Using
$p_j^{**}=p_j$, we may put $d=e^{i\phi}$, $g_j=ie^{i\phi/2}c_j$,
$\phi,c_j\in\rb$ ($j\in\spi$). Setting $m=e^{i\phi/2}$, $n_j=\frac12ic_j$, 
one
gets that ${p\pri}_i$ given by \r{3.4a} satisfy ${p\pri}_i^*={p\pri}_i$. 

We put $c=\frac1m$, $h_j=-\frac1{m}n_j$, $M=\jb$, $\hat\api=\api$,
$\phi_{\api}=\id$. Then 
$\hat\bpi,\hat\api,\hat\de,\hat\la,\hat p,\hat f,\hat\eta,\hat T$ and 
$\phi:\bpi\lra\hat\bpi$ given by Proposition \re{p3.4}.3 satisfy all the
conditions which don't involve $*$. In particular, $\la=\hat\la$, 
\[ \phi({p\pri})=\phi(mp+n-\la n)=m(c\hat p+h-\la h)+n-\la n=\hat p. \]
We define $*$ in $\hat\bpi$ as $\phi\circ *_{\bpi}\circ\phi^{-1}$. Then
$\phi$ is a Hopf ${}^*$-algebra isomorphism and 
$\hat p_i^*=\phi({p\pri}_i^*)=\phi({p\pri}_i)=\hat p_i$. 

ad 2) The existence of such structure is equivalent to the fact that the 
ideal
${\cal G}$ in $\te\bpi$ (with $*$ given by $\te p_i^*=\te p_i$,
$*_{|_{\api}}=*_{\api}$) generated by \r{2.4} and \r{2.64} is selfadjoint 
(Hopf algebra structure exists due to Proposition \re{p2.14H}, while
$\de*=(*\ot*)\de$ and $*^2=\id$ can be checked on generators $a\in\api$, $p_i$
($i\in\spi$)). In
other words, conjugating \r{2.4} and \r{2.64} we should get relations, which
follow from \r{2.4} and \r{2.64} as relations in algebra without $*$. We use
the notation $\bar f(x)=\ov{f(x^*)}$, $f\in\api'$, $x\in\api$. For
\r{2.4} one gets ($b=a^*$)
\[ 0=[-p_sa+(a*f_{st})p_t+\phi_s(a)]^*=
-bp_s+p_t(b*\ov{f_{st}})+\phi_s(b^*)^*= \]
\[ -bp_s+(b*\ov{f_{st}}*f_{tr})p_r+\phi_t(b*\ov{f_{st}})+\phi_s(b^*)^*. \]
Therefore (see \r{2.2a})
$\ov{f_{st}}*f_{tr}=\del_{sr}\eps$, 
$\ov{f_{sl}}=\ov{f_{st}}*f_{tr}*f_{rl}\circ S=f_{sl}\circ S$, 
$f_{sl}=\ov{f_{sl}\circ S}$, we get \r{3.5'}. Moreover, 
\be \phi_s(b^*)=-\phi_t(b*\ov{f_{st}})^*.\e{3.7'}\ee
Thus
\[ b^**\eta_s-\la_{st}(\eta_t*b^*)=
-b^**f_{st}*\ov{\eta_t}+(\ov{\eta_m}*b^**f_{st})\la_{tm}= \]
\[ -b^**f_{st}*\ov{\eta_t}+\la_{st}(f_{tm}*\ov{\eta_m}*b^*) \]
(we used \r{2.3}). Thus ($b^*=a$) $a*\mu_s=\la_{st}(\mu_t*a)$, where
$\mu_s=\eta_s+f_{st}*\ov{\eta_t}$. Inserting $a=v_{kl}$, $v\in \Irr\ H$, and
denoting $\mu_s(v_{km})=F_{sk,m}$, one has $F\in\mor(v,\la\tp v)=\{0\}$
(condition c.), $\mu_s=0$. Therefore (see \r{2.2a}), 
\[ -f_{ms}\circ S*\eta_s=f_{ms}\circ S*f_{st}*\ov{\eta_t}=\ov{\eta_m}.
\]
Acting on $v_{kl}$ and using \r{2.15}, we obtain 
\[ \ov{\eta_m}(v_{kl})=-f_{ms}(v_{kr}^{-1})\eta_s(v_{rl})=
-\eta_m(\del_{kl}I)+\eta_m(v_{kr}^{-1})\eps(v_{rl})=\eta_m(v_{kl}^{-1}), \]
$\ov{\eta_m}=\eta_m\circ S$, one gets \r{3.6'}. 

In virtue of \r{3.5'}, \r{2.2a} and $\bar\la=\la$
\[ \del_{ij}\del_{kl}=\del_{ij}\eps(\la_{kl})=f_{im}*\ov{f_{mj}}(\la_{kl})= \]
\[ f_{im}(\la_{kr})\ov{f_{mj}}(\la_{rl})=R_{ik,rm}\ov{R_{mr,lj}}. \]
Multiplying by $R_{ab,ik}$, we get
\be R_{ab,jl}=\ov{R_{ba,lj}}.\e{3.8'}\ee
Moreover,
\[ \eta_s(\la_{jk})=-(f_{st}*\ov{\eta_t})(\la_{jk})=
-f_{st}(\la_{jm})\ov{\eta_t}(\la_{mk})=-R_{sj,mt}\ov{\eta_t(\la_{mk})}. \]
Multiplying by $(R-\jb^{\ot 2})_{ab,sj}$, one obtains 
\be (R-\jb^{\ot 2})_{ab,sj}\eta_s(\la_{jk})=
(R-\jb^{\ot 2})_{ab,mt}\ov{\eta_t(\la_{mk})}.\e{3.9}\ee
Therefore, conjugating \r{2.64}, we get
\[ (R-\jb^{\ot 2})_{lk,ji}[p_jp_i-\eta_j(\la_{is})p_s+\te T_{ji}-
\la_{jn}\la_{im}\te T_{nm}]=0. \]
Comparing with \r{2.64}, one has
$(R-\jb^{\ot 2})(\te T-T)\in\mor(I,\la\tp\la)$. Using \r{2.67} and 
\r{3.8'}, we
get \r{3.6a}. Conversely, assuming \r{3.5'}--\r{3.6a} and repeating the above
reasonings in an opposite order, one gets that ${\cal G}$ is selfadjoint. 

Since $\api$ and $p_i$ ($i\in\spi$) generate $\bpi$, uniqueness of $*$
follows. 

ad 3) If $\phi\circ *=*\circ\phi$ and $\bar M=M$ then (in 2) of Proposition
\re{p3.4}) 
\[ M\ov{\hat\la}M^{-1}=\ov{\phi(\la)}=\phi(\bar\la)=
\phi(\la)=M\hat\la M^{-1},
\]
hence $\ov{\hat\la}=\hat\la$. Moreover, 
\[ M(\bar c\hat p+\bar h-\hat\la\bar h)=\ov{\phi(p)}=\phi(\bar p)=\phi(p)= 
M(c\hat p+h-\hat\la h). \]
Thus $\bar h-h\in\mor(I,\hat\la)=\{0\}$ (condition c.), $c,h_i\in\rb$
($i\in\spi$). In 3) of Proposition \re{p3.4} we prove $\ov{\hat\la}=\hat\la$
as above and 
define Hopf ${}^*$-algebra 
structure
in $\hat\bpi$ by $*_{\hat\bpi}=\phi\circ *_{\bpi}\circ\phi^{-1}$, which gives
the proper $*$ in $\hat\api$. Thus $\phi$ is a 
Hopf ${}^*$-algebra isomorphism by
construction. Then 
\[ M(c\ov{\hat p}+h-\hat\la h)=\ov{\phi(p)}=\phi(\bar p)=\phi(p)=
M(c\hat p+h-\hat\la h), \]
$\ov{\hat p}=\hat p$.\ed

\bp\e{p3.6} Let $\bpi$ satisfy the conditions of Proposition \re{p3.5}.2 with
$\te T=T$. Then (see \r{2.68})
$\tau^{kl}(S(b^*))=\ov{\tau^{lk}(b)}$.\ep

\bd Using $\de *=(*\ot *)\de$, $\de S=\tau(S\ot S)\de$, $\eps S=\eps$, 
\r{3.5'},
\r{3.6'} and $T_{ij}=\ov{T_{ji}}$ one gets 
\[ \tau_{ij}(S(b^*))=\ov{\eta_i*\eta_j(b)}-
\eta_i(\la_{js})\ov{\eta_s(b)}+\ov{T_{ji}\eps(b)}-
\ov{(f_{im}*f_{jn})(b)T_{nm}}. \]
Multiplying both sides by $(R-\jb^{\ot 2})_{kl,ij}$ and using \r{3.8'},
\r{3.9}, one gets our assertion.\ed

\bc\e{c3.7} With assumptions of Proposition \re{p3.6}, if \r{2.69} holds for
some $b\in\api$ then \r{2.69} holds for $S(b^*)$.\ec

\bd Let $c=S(b^*)$. Applying $S\circ *$ to \r{2.69} and using
Proposition \re{p3.6}, one obtains 
$(\la^{-1})_{ki}(\la^{-1})_{lj}(c*\tau^{ji})=\tau^{lk}*c$, which is 
equivalent
to \r{2.69} with $b$ replaced by $c$.\ed

\sectio{Quantum homogeneous spaces}

In this Section we prove that each inhomogeneous quantum group possesses
(under some conditions) exactly one analogue of Minkowski space.
Throughout the Section we assume that $\po(H)=(\api,\de)$ is a Hopf 
${}^*$-algebra
satisfying the conditions a.--c. and $\po(G)=(\bpi,\de)$ is the corresponding
Hopf ${}^*$-algebra 
as in Theorems \re{t2.5}, \re{t2.8} with $*$-structure as in
Proposition \re{p3.5}.2. 

\br Analogues of Minkowski spaces endowed with the action of inhomogeneous 
quantum group in absence of inhomogeneous terms in the commutation relations
were studied e.g. in \cite{S}, \cite{M}, \cite{ChD}, 
for the so called soft deformations
(a commutative $\api$ and $\eta=0$) in \cite{GQ} and for $\kappa$-Poincar\'e
group in \cite{L}.\er

Motivated by \cite{QP} we say that $(\cpi,\Psi)$
describes an analogue of Minkowski space associated with $G$ if one has

\btr \item[1.] $\cpi$ is a unital 
${}^*$-algebra generated by $x_i$, $i\in\spi$,
and $\Psi :\cpi\lra \bpi\ot\cpi$ is a unital ${}^*$-homomorphism such that
$(\eps\ot\id)\Psi=\id$, $(\id\ot\Psi)\Psi=(\de\ot\id)\Psi$, 
$x_i^*=x_i$ and
\be \Psi x_i=\la_{ij}\ot x_j+p_i\ot I.\e{3a.1}\ee
\item[2.] if $\Psi W\subset \api\ot W$ 
for a linear subspace $W\subset\cpi$ then
$W\subset\cb I$.
\item[3.] if $(\cpi',\Psi')$ also satisfies 1.--2. for some 
${x_i}'\in\cpi'$ then
there exists a unital ${}^*$-homomorphism $\rho:\cpi\lra\cpi'$ such that
$\rho(x_i)={x_i}'$ and $(\id\ot\rho)\Psi=\Psi'\rho$ (universality of 
$(\cpi,\Psi)$).
\etr

Let us remark that the conditions 
$(\eps\ot\id)\Psi=\id$, $(\id\ot\Psi)\Psi=(\de\ot\id)\Psi$
are superfluous in 1..

\bp\e{p3a.1} We assume 
\be \mor(I,\la\tp\la)\cap\ker(R+\jb^{\ot 2})=\{0\}.\e{3a.0}\ee
Let $\cpi'$ be a unital algebra generated by $x_i$ ($i\in\spi$) and
$\Psi':\cpi'\lra\bpi\ot\cpi'$ be a unital homomorphism satisfying \r{3a.1} and
the condition 2.. Then 
\be (R-\jb^{\ot 2})_{ij,kl}(x_k x_l-\eta_k(\la_{lm})x_m+T_{kl})
=0.\e{1.17}\ee\ep

\bd According to \r{3a.1}, $\Psi' x=\la\ti x+p\ti I$. Thus
\[ \Psi'(x\tp x)=(\la\tp\la)\ti(x\tp x)+(\la\tp p)\ti x+(p\tp\la)\ti x+
(p\tp p)\ti I. \]
Using \r{2.17a}, \r{2.69'} and \r{2.68'},
\[ \Psi'((R-\jb^{\ot 2})(x\tp x))=(\la\tp\la)\ti(R-\jb^{\ot 2})(x\tp x)+ \]
\[ (R-\jb^{\ot 2})(Z\la-(\la\tp\la)Z)\ti x+
(R-\jb^{\ot 2})(Zp-T+(\la\tp\la)T)\ti I. \]
Setting $w=(R-\jb^{\ot 2})(x\tp x-Zx+T)$, one obtains 
$\Psi' w=(\la\tp\la)\ti w$. The condition 2. implies $w_{ij}\in\cb I$, 
$w=(\la\tp\la)w$. But $Rw=-w$, hence 
$w\in\mor(I,\la\tp\la)\cap\ker(R+\jb^{\ot 2})=\{0\}$ and \r{1.17} 
follows.\ed 

\bp\e{p3a.2} Assume that 
\be \mor(I,\la\tp\la\tp\la)=\{0\}.\e{3a.2}\ee
Let $\cpi$ be the 
universal unital algebra generated by $x_i$ ($i\in\spi$) satisfying
\r{1.17}. Then there exists a unique unital homomorphism
$\Psi:\cpi\lra\bpi\ot\cpi$ such that \r{3a.1} holds. Moreover, 
$\al^{in}(x^{\tu n})$, $i=1,2,\k,\dim S_n$, $n=0,1,\k,N$, form a basis of 
\[ \cpi^N=\spa\{x_{i_1}\cdot\k\cdot x_{i_n}:
\ i_1,\k,i_n\in\spi,\ n=0,1,\k,N\}, \] 
and the condition 2. holds. In particular,
\[ \dim\cpi^N=\sum^N_{n=0} \dim S_n.  \]\ep

\bd Doing the same computations as in the proof of
Proposition \re{p3a.1}, 
we get that
the right hand sides of \r{3a.1} satisfy \r{1.17} in
$\bpi\ot\cpi$. Therefore the desired $\Psi$ exists. Uniqueness is trivial.
We find the basis of $\cpi^N$ in a similar way as the basis of the
left module ${\bpi\pri}^N$ in 
Section 3 (Theorem \re{t2.8}--Corollary \re{c2.9}).
The main change is that we now consider the equality $(x\tp x)\tp x=x\tp(x\tp
x)$, using $x\tp x=R(x\tp x)+k$, where $k=c\cdot x+(R-\jb^{\ot 2})T$. One 
gets
$A(k\tp x)=B(x\tp k)$. Instead of \r{2.49}--\r{2.50} we have 
\be l=-\frac12 Hc,\e{3a.3}\ee
\be 0=-\frac12 H(R-\jb^{\ot 2})T\e{3a.4}\ee
where
\[ l=A((R-\jb^{\ot 2})T\ot\jb)-B(\jb\ot(R-\jb^{\ot 2})T)=
-A_3(T\ot\jb-\jb\ot T)=L. \]
Thus \r{3a.3} is equivalent to \r{2.49}, which is equivalent to \r{2.31b}.
Moreover, using \r{2.67}, \r{2.54} and \r{3a.2}, one obtains that \r{3a.4} is
equivalent to \r{2.32}. Then we find the basis in a similar way as in Section
3. 

Now we shall prove the condition 2.. Let $y\in\cpi$, $\Psi y\in\api\ot\cpi$. If
$y\not\in\cb I$ then for some $N>0$ 
\[ y=\sum_{i=1}^{\dim\ S_N} c_i\al^{iN}x^{\tu N}+y', \]
where $y'\in\cpi^{N-1}$ and not all $c_i$ equal $0$. Using \r{3a.1},
\[ \Psi y=\sum_{i=1}^{\dim\ S_N}(c_i\al^{iN}p^{\tu N})\ot I+\omega, \]
where $\omega\in\bpi^{N-1}\ot\cpi$ and also $\Psi y\in\bpi^{N-1}\ot\cpi$. In
virtue of Corollary \re{c2.11} $c_i=0$. This contradiction shows that
$y\in\cb I$.\ed

\br\e{r3a.2'} Assume \r{3a.0} and \r{3a.2}. Then
 $(\cpi,\Psi)$ defined in 
Proposition~\re{p3a.2} satisfies the 
conditions 1.--3. without $*$ (one can check $(\eps\ot\id)\Psi=\id$, 
$(\id\ot\Psi)\Psi=(\de\ot\id)\Psi$, $(\id\ot\rho)\Psi=\Psi'\rho$ on 
generators).
The existence of $*$-structures in $\api$, $\bpi$ is not necessary for
Proposition~\re{p3a.1}, Proposition~\re{p3a.2} and Remark~\re{r3a.2'}.\er

\bp\e{p3a.3} Let \r{3a.0} and \r{3a.2} hold and $(\cpi,\Psi)$ 
be as in Proposition
\re{p3a.2}.
Then there exists a unique ${}^*$-algebra 
structure in $\cpi$ such that $\Psi$ 
is a
${}^*$-homomorphism. It is determined by $x_i^*=x_i$.\ep

\bd Assume that $\cpi$ is a ${}^*$-algebra and $\Psi$ is a 
${}^*$-homomorphism.
Conjugating \r{3a.1} and comparing with \r{3a.1}, one gets 
$\Psi z_i=\la_{ij}\ot z_j$ where $z_i=x_i^*-x_i$. Using the condition 2. 
(see
Proposition \re{p3a.2}), $z_i=k_iI$ with $k_i\in\cb$. Thus 
$k=(k_i)_{i=0}^3\in\mor(I,\la)=\{0\}$, $z_i=0$, $x_i^*=x_i$. Since we must
have $I^*=I$, it determines $*$ in $\cpi$ uniquely. 

Conversely, setting $x_i^*=x_i$ in free unital algebra generated by $x_i$, we
get a ${}^*$-algebra. In virtue of \r{3.6a}, \r{3.8'} and \r{2.67}
\[ \te T-T\in\mor(I,\la\tp\la)\cap\ker(R+\jb^{\ot 2})=\{0\}. \]
Using this, \r{3.8'} and \r{3.9}, one checks that the ideal generated by the
left hand sides of \r{1.17} is selfadjoint. Hence there exists ${}^*$-algebra
structure in $\cpi$ such that $x_i^*=x_i$. Using \r{3a.1}, 
$\Psi\circ *=*\circ\Psi$ on $x_i$, hence in whole $\cpi$, and $\Psi$ is a
${}^*$-homomorphism.\ed 

\bt\e{t3a.5} Assume \r{3a.0} and \r{3a.2}. Then the
conditions 1.--3. are satisfied if and only if the pair $(\cpi,\Psi)$ is
$*$-isomorphic to that defined in Propositions \re{p3a.2} and \re{p3a.3}.\et

\bd According to Propositions \re{p3a.2} and \re{p3a.3}, $(\cpi,\Psi)$ 
satisfies
the conditions 1.--2.. If $(\cpi',\Psi')$ also satisfies 1.--2., then using
Proposition \re{p3a.1}, \r{1.17} is satisfied in $\cpi'$ and there exists a
unital homomorphism $\rho:\cpi\lra\cpi'$ such that $\rho(x_i)={x_i}\pri$.
Using $x_i^*=x_i$, ${{x_i}\pri}^*={x_i}\pri$, one gets that 
$\rho\circ *=*\circ\rho$. Using \r{3a.1}, one gets $(\id\ot\rho)\Psi=
\Psi'\rho$ on
$x_i$ and hence in whole $\cpi$. Thus the condition 3. is satisfied.
Uniqueness follows from the universality.\ed

\bp\e{p3a.6} Assume \r{3a.0} and \r{3a.2}. Let $\phi:\bpi\lra\hat\bpi$ be a
Hopf ${}^*$-algebra isomorphism of quantum inhomogeneous groups, $p_i^*=p_i$, 
$\hat p_i^*=\hat p_i$, $\phi_{|_{\api}}=\phi_{\api}:\api\lra\hat\api$
be a Hopf ${}^*$-algebra isomorphism such that 
$\phi_{\api}(\la)=M\hat\la M^{-1}$, $\phi(p)=M(c\hat p+h-\hat\la h)$, 
$\bar M=M$, $c,h_s\in\rb$ ($s\in\spi$). Let $(\cpi,\Psi)$ and 
$(\hat\cpi,\hat\Psi)$ be
the corresponding objects satisfying 1.--3.. Then there exists 
a unital ${}^*$-isomorphism
$\phi_{\cpi}:\cpi\lra\hat\cpi$ such that $\phi_{\cpi}(x)=M(c\hat x+h)$ and
$\phi, \phi_{\cpi}$ intertwine $\Psi$ with $\hat\Psi$.\ep

\bd We set $\te x=c^{-1}(M^{-1}x-h)$ and check that $(\cpi,(\phi\ot\id)\Psi)$
satisfies the conditions 1.--3. w.r.t. $\te x$, $\hat\la$ 
and $\hat p$. By virtue of
universality, there exists a unital ${}^*$-isomorphism
$\phi_{\cpi}:\cpi\lra\hat\cpi$ such that $\phi_{\cpi}(\te x)=\hat x$, 
$(\phi\ot\phi_{\cpi})\Psi=\hat\Psi\phi_{\cpi}$.\ed\newpage

\begin{center} {\bf Acknowledgements} \end{center}

The first author is grateful to Prof. W. Arveson and other faculty members
for their kind hospitality in UC Berkeley.  
The authors are thankful to Prof.~J.~Lukierski and Dr~S.~Zakrzewski for
fruitful discussions.\\\\

\mbox{}\\
{\bf Note}. In Ref. 11 one should assume that 
${\cal A}_0$ has an invertible coinverse. 
\end{document}